\documentclass{aa}  

\usepackage{graphicx}
\usepackage{txfonts}
\usepackage{lipsum}
\usepackage{subcaption}         
\usepackage{lscape}             
\usepackage{placeins}           

\usepackage{color} 

\def\eimt{EIM-tool}

\begin{document}

\title{Constraining the shape of dark matter haloes using only starlight}
\subtitle{II. Tests of the technique with objects of known gravitational potential}

\author{Jorge S\'anchez Almeida\inst{1,2}
  \and
  Angel R. Plastino\inst{3}
  \and
  Sergio Guerra Arencibia\inst{1,2}
  \and \\
  Nitya Kallivayalil\inst{4}
  \and
   Jack T. Warfield\inst{4}
       }
       \institute{
       {Instituto de Astrof\'\i sica de Canarias, La Laguna, Tenerife, E-38200, Spain}\\
       \email{jos@iac.es,sergio.guerra@iac.es}
       \and
       Departamento de Astrof\'\i sica, Universidad de La Laguna
       \and
       {CeBio y Departamento de Ciencias B\'asicas, 
       Universidad Nacional del Noroeste de la Prov. de Buenos Aires, 
       UNNOBA, CONICET, Roque Saenz Pe\~na 456, Junin, Argentina}\\
     \email{arplastino@unnoba.edu.ar}
     \and
     {Department of Astronomy, University of Virginia, 530 McCormick Road, Charlottesville, VA 22904 USA}\\
     \email{njk3r@virginia.edu,jtw5zc@virginia.edu}
       }

   \date{Received \today; accepted \dots}


   \abstract{
Under the collisionless cold dark matter (CDM) paradigm, galaxies with stellar masses below $10^{5-6} M_\odot$ are expected to preserve primordial cuspy dark matter (DM) profiles. Because baryonic feedback should be too weak to transform cusps into cores at these masses, such galaxies provide especially sensitive tests of DM physics. If cores are observed in these systems, they could indicate departures from CDM. To address this problem, \citet{2025A&A...694A.283S} introduced the Eddington Inversion Method tool (\eimt), a photometry-based diagnostic that avoids the need for long-term spectroscopic observations and is suitable for the analysis of the forthcoming large photometric surveys of galaxies. The method relies on the fact that embedding a stellar distribution with a central core inside a cuspy NFW halo can require a negative phase-space distribution function $f$, which is physically impossible. By fitting observed stellar surface densities using $f$ as a free parameter,  \eimt\ tests whether an assumed gravitational potential is physically consistent with an observed stellar distribution.
This paper validates \eimt\ using globular clusters, dwarf spheroidal galaxies (dSphs), and numerical simulations with independently known potentials. Applied to 21 globular clusters, the method rejected NFW cusps in 71\,\% of cases while consistently favoring cored Schuster–Plummer models, accurately recovering expected core radii. Applied to dSphs, \eimt\ reproduced conclusions from classical dynamical studies: Sculptor and Fornax favored cored halos, whereas Draco remained compatible with a cusp. Tests on EDGE and FIRE simulations further showed that the tool reliably rejects cuspy NFW profiles when the true DM distribution is core-like. Overall, these results establish \eimt\ as a robust and efficient photometric method for distinguishing cored and cuspy DM halos.
 }


\titlerunning{Constraining the shape of dark matter haloes using only starlight II}
\authorrunning{J. S\'anchez Almeida et al.}

\maketitle


        %
\begin{table*}
\caption{Tests  on \eimt\ carried out in this work}              
\label{table:summary}      
\centering                                      
\begin{tabular}{p{4cm}p{4cm}p{4cm}c}          
\hline\hline                        
  \noalign{\smallskip}
  \centering Idea & \centering Actual Test & \centering Result & Passed? \\    
  \noalign{\smallskip}
  \hline
  \noalign{\smallskip}
  \centering The stars in Globular Clusters (GCs) are self-gravitating. The distribution of stars and potential must agree.
                  & \centering (1) Apply \eimt\ to 21 GCs from \citet{2013ApJ...774..151M} -- Sect.~\ref{sec:globular}& GCs have cores. \eimt\ favors the stars to reside in core potentials rather than in NFW potentials. Core size inferred for potentials similar to core size for GCs.  -- Figs.~\ref{fig:df6_run_plot_111} and \ref{fig:df6_run_plotg}.&Yes\\
  \noalign{\smallskip}
  \hline
  \noalign{\smallskip}
\centering Dwarf galaxies with gravitational potential known through kinematic analysis & \centering (2) Sculptor. Data and Jeans analysis from {\catcode \&=12 \citet{2025A&A...699A.347A}} -- Sect.~\ref{sec:sculptor}&Sculptor DM halo is is not NFW, as inferred from Jeans analysis -- Figs.~\ref{fig:df6_run_plot_1_pub} and \ref{fig:df6_run_plotd}& Yes\\
  \noalign{\smallskip}
   &\centering (3) Draco. Data from Gaia (Appendix~\ref{app:draco}) -- Sect.~\ref{sec:draco} & NFW allowed, in agreement with previous works -- Figs.~\ref{fig:df6_run_plot_13_pub} and \ref{fig:df6_run_ploth_battaglia1}&Yes\\
  \noalign{\smallskip}
& \centering (4) Fornax dSph.  Data from \citet{Guerra+26} -- Sect.~\ref{sec:fornax} & Extended inner plateau, incompatible with NFW, in agreement with previous works -- Figs.~\ref{fig:df6_run_plot_531_pub} and \ref{fig:df6_run_plotd_fornax_sguerra3}.&Yes\\
  \noalign{\smallskip}
 \hline
  \noalign{\smallskip}
\centering Numerical simulations provide model galaxies of known DM distribution& \centering (5) EDGE halos -- Sect.~\ref{sec:edge}& \centering NFW halos are discarded, in agreement with the true DM distribution -- Fig.~\ref{fig:df6_run_plot_411_pub} &Yes\\ 
  \noalign{\smallskip}
&\centering (6) FIRE halo -- Appendix~\ref{app:fire}&\centering NFW halo is discarded, in agreement with the true DM distribution -- Fig.~\ref{fig:df6_run_plot_601_pub} &Yes\\ 
  \noalign{\smallskip}
  \hline
\hline                                             
\end{tabular}
\end{table*}

\section{Introduction}\label{sec:intro}

The nature of dark matter (DM) remains one of the fundamental open questions in modern science. The standard cosmological model treats DM as cold and collisionless particles (CDM), interacting with baryonic matter only through gravity. While this framework has proven remarkably successful, it is likely an effective theory that conceals a deeper, more complex physical reality. Due to the minimal interaction with baryons, any direct study of DM is extremely difficult. Therefore, the analysis of observational tensions between CDM predictions and real astronomical data represents a primary window into the true nature of DM and the physics beyond the current approximation \citep[e.g.,][]{2021arXiv210602672P}.

The core-cusp problem represents one of such possible tensions. CDM simulations predict steep central density cusps in DM halos, whereas observations of many galaxies reveal shallower, constant-density cores \citep[e.g.,][]{1997ApJ...490..493N,2004MNRAS.351..903G,2017Galax...5...17D,2017ARA&A..55..343B,2019A&ARv..27....2S}. Baryonic feedback from star formation can account for this discrepancy by redistributing DM \citep{2010Natur.463..203G,2012MNRAS.421.3464P}.  However, this mechanism becomes ineffective below a stellar mass threshold of $\sim 10^{5-6}\,M_\odot$, meaning that galaxies with lower masses are expected to retain their primordial cusps under CDM \citep[e.g.,][]{2012ApJ...759L..42P,2016MNRAS.456.3542T,2020MNRAS.497.2393L}. Thus, objects with masses below this threshold possess strong diagnostic potential, since the presence of DM cores in such systems would likely indicate deviations from the CDM approximation. They therefore provide a gateway to investigating the true nature of DM.

Determining whether such extremely faint galaxies possess DM cores is challenging. The traditional approach relies on multi-epoch spectroscopy to measure stellar velocities while accounting for effects such as the presence of binary stars \citep[e.g.,][]{2022ApJ...939....3P,2023A&A...677A..95A,2024ApJ...970....1V}. As a result, this method is observationally very expensive. We recently developed a new technique, based solely on photometry, that allows us to determine whether the observed stellar distribution in a galaxy is inconsistent with the cuspy DM halos predicted by CDM.
Under a set of reasonable assumptions (such as spherical symmetry and absence of tangential velocity bias), the Eddington inversion method shows that the phase-space distribution function ($f$) required for a stellar core embedded in a cuspy CDM potential is negative, highlighting the physical inconsistency between the two \citep[][]{1992MNRAS.255..561C,2006ApJ...642..752A,2023ApJ...954..153S,2024A&A...690A.151S}.
The inconsistency holds not only for spherically symmetric stellar systems with isotropic velocities, but also for axisymmetric systems \citep{2024A&A...690A.151S}, triaxial systems (Plastino et al., in preparation), radially biased orbits (J. Sánchez Almeida et al. 2023), and even potentials with small cores \citep{1992MNRAS.255..561C,2024RNAAS...8..167S}. Motivated by the idea of exploiting the {\em core in cusp} inconsistency for diagnostics, we developed a tool to fit the observed stellar light distribution in a galaxy, treating the function $f$ as a free parameter. In this framework, if the inferred $f$ becomes negative in any region of the phase space, it signals an incompatibility with the assumed gravitational potential. This approach is implemented in the method described by \citet[][hereinafter \eimt]{2025A&A...694A.283S}. When applied to the surface density profiles of the six ultra-faint dwarf galaxies (UFDs) observed by \citet{2024ApJ...967...72R}, it indicates that these systems are inconsistent with CDM potentials \citep{2024ApJ...973L..15S}. Since stellar feedback is expected to be negligible in UFDs, the most compelling interpretation is that their DM departs from the standard CDM paradigm, showing the diagnostic power of \eimt .

Paper~I \citep{2025A&A...694A.283S}, where \eimt\ was introduced, describes a number of sanity checks, among them the analysis of mock stellar mass surface density profiles with added noise consistent with the assumed potential. The second paper in this series aims at a more comprehensive validation of the tool using both observed systems and numerical simulations with known gravitational potentials. The tests consist of comparing the outcome of \eimt\ with the known properties of the potentials, obtained independently. Globular clusters (GCs) are ideal testbeds. They are self-gravitating systems in which light traces mass, so that their cored surface density is incompatible with cuspy potentials, providing a specific prediction for testing. We also use a number of dwarf spheroidal galaxies (dSphs) whose gravitational potentials are known from estimates based on classical dynamical methods.  Finally, we test \eimt\ on a number of model galaxies drawn from numerical simulations in which the DM distribution is computed self-consistently. The actual list of tests is indicated in Table~\ref{table:summary}.

The paper is organized as follows:
Section~\ref{sec:briefing} provides a general description of \eimt\ and the procedure used to test it.
Section~\ref{sec:globular} applies the tool to globular clusters (GCs), known to contain little, if any, DM, and thus characterized by a potential created by their cored stellar distribution.
Section~\ref{sec:dwarfs} treats with \eimt\ three dwarf spheroidal galaxies with measured potential. In particular, Draco (Sect.~\ref{sec:draco}) was chosen because existing analyses make its potential consistent with a cusp.
Section~\ref{sec:edge} applies the tool to numerical simulations where the mass distribution responsible for the potential is known. It uses EDGE galaxies \citep{2020MNRAS.491.1656A}, a result complemented in Appendix~\ref{app:fire} with a FIRE halo \citep{2018MNRAS.480..800H}.
Section~\ref{sec:discussion} gives a summary with the conclusions.

%
%
\section{Briefing on \eimt}\label{sec:briefing}

For completeness and to clarify the notation, we provide here a summary of \eimt. Full details can be found in Paper~I \citep{2025A&A...694A.283S}, to which we refer the reader for further clarification.

In a spherically symmetric system of particles with isotropic velocity distribution, the phase-space distribution function, $f$, depends only on the particle energy $\epsilon$. Then, the volume density $\rho(r)$ can be written as  \citep[e.g.,][]{2008gady.book.....B}
\begin{equation}
  \rho(r) = \int_0^{\epsilon_{max}}f(\epsilon)\,\xi(\epsilon,r)\, d\epsilon,
  \label{eq:master}
\end{equation}
with $\epsilon = \Psi - \frac{1}{2} v^2$ the relative energy per unit mass of each  particle, $v$ the particle's velocity, and $\Psi(r) = \Phi_0 - \Phi(r)$ the relative potential energy at a distance $r$ from the center of the system.  The symbol $\Phi(r)$ stands for the gravitational potential energy  and $\Phi_0$ is the gravitational potential energy evaluated at the edge of the system. The kernel $\xi(\epsilon,r)$ stands for
\begin{equation}
  \xi(\epsilon,r)=4 \pi \sqrt{2}\sqrt{\epsilon_{\rm max}}\sqrt{\frac{\Psi(r)}{\Psi(0)}-\frac{\epsilon}{\epsilon_{max}}} \,\,\Pi(X-r),
  \label{eq:master_mind}
\end{equation}
where $\epsilon_{max}= \Psi(0)$, $X$ is a radius set by $\epsilon$ and implicitly defined by the relation $\Psi(X)/\Psi(0) = \epsilon/\epsilon_{max}$, and $\Pi(x)$ represents the step function,
\def\cacab{if $x < 0$}
\def\cacac{if $x \geq 0$}
\begin{equation}
  \Pi(x) =  
  \begin{cases}
   0 & {\rm if~} x < 0 ,\\
   1 & {\rm if~} x \geq 0.
   \end{cases}
 \end{equation}
 Equation~(\ref{eq:master}) can be interpreted in a way that highlights its physical significance. The kernel $\xi(\epsilon,r)$ parameterizes a family of energy dependent volume densities characteristic of the potential $\Psi$. Then, $\rho(r)$ is just the superposition of these  characteristic densities with $f(\epsilon)$ quantifying the contribution of each energy (see Eq.~[\ref{eq:master}]).
In principle, $f(\epsilon)$ could be retrieved using Eq.~(\ref{eq:master}) by fitting $\rho(r)$ with a linear superposition of $\xi(\epsilon,r)$. In practice, however, there is no unique way to discretize Eq.~(\ref{eq:master}).  Keeping in mind that any well behaved function admits a polynomial approximation, we approach the practical problem  expanding $f(\epsilon)$ as a polynomial of order $n$, 
\begin{equation}
  f(\epsilon)\simeq \sum_{i=3}^n\,\frac{a_{i}}{\epsilon_{max}^{3/2}}\,(\epsilon/\epsilon_{max})^i,
  \label{eq:polydef}
\end{equation}
so that
\begin{equation}
  \rho(r) \simeq \sum_{i=3}^n\,a_i\,F_i(r),
\end{equation}
with
\begin{equation}
  F_i(r) =\int_0^{1}\,\alpha^i\,\frac{\xi(\alpha\,\epsilon_{max},r)}{\sqrt{\epsilon_{max}}}\,d\alpha .
  \label{eq:master2}
\end{equation}
We note that the polynomial expansion in Eq.~(\ref{eq:polydef}) lacks the three first terms (it begins at $i=3$). This is a constraint imposed by the need to have a finite total mass \citep[see][]{2025A&A...694A.283S}. The discretization also holds for the projection of the 3D densities in the plane of the sky, i.e., 
\begin{equation}
 \Sigma(R) \simeq \sum_{i=3}^n\,a_i\, S_i(R),
  \label{eq:master3}
\end{equation}
\begin{equation}
 S_i(R) = \int_0^{1}\,\alpha^i\,\frac{\xi_\Sigma(\alpha\epsilon_{max},R)}{\sqrt{\epsilon_{max}}}\,d\alpha,
  \label{eq:master4}
\end{equation}
where $\Sigma(R)$ and $\xi_\Sigma(\epsilon,R)$ are the 2D projection (the Abel transform) of $\rho(r)$ and $\xi(\epsilon,r)$, respectively. The symbol $R$ stands for the radial coordinate in the plane of the sky projection.
Except for the arbitrary scaling given by $\epsilon_{max}$, Eqs.~(\ref{eq:polydef}) and (\ref{eq:master3}) provide a method to infer the $f(\epsilon)$ needed to account for the stellar surface density $\Sigma(R)$ provided the stars reside in an assumed  gravitational potential. A fitting algorithm using  Eq.~(\ref{eq:master3}) yields the coefficients $a_i$ determining $f(\epsilon)$ through Eq.~(\ref{eq:polydef}). The kernel characteristic of the potentials has to be computed numerically. It is a cumbersome task, so \citet{2025A&A...694A.283S} evaluated them for only three potentials: the NFW potential (after Navarro, Frenk, and White) was chosen because it provides the canonical description of CDM halos \citep{1997ApJ...490..493N}. The Schuster–Plummer potential is a polytrope and therefore represents a self-gravitating system in a state of maximum entropy \citep{1993PhLA..174..384P}. Polytropes reproduce remarkably well both the dark matter distribution observed in dwarf galaxies \citep{2020A&A...642L..14S} and their stellar distribution \citep{2021ApJ...921..125S}. The Schuster–Plummer model corresponds to the polytrope of index 5, but polytropes of any index share the same central behavior within the cored region \citep[e.g.,][]{2022Univ....8..214S}. Consequently, the Schuster–Plummer potential can be regarded as representative of the entire family of polytropic potentials. Finally, the potential arising from a DM distribution following a $\rho_{230}$ profile,
\begin{equation}
\rho_{230}(r) = \frac{\rho_s}{\left[1+(r/r_s)^2\right]^{3/2}},
\end{equation}
was chosen because it has a central core similar to Schuster-Plummer and a tail similar to NFW. The symbols $\rho_s$ and $r_s$ represent two arbitrary constants.

\eimt\ implements the above equations to infer $f(\epsilon)$ by fitting $\Sigma(R)$. The free parameters are the amplitudes $a_i$ together with a global radial scaling factor that sets the width of the potential; the presence of this scaling renders the fit non-linear.
The fits were carried out using a Bayesian approach, with the log-likelihood defined as $-\chi^2/2$ where
\begin{equation}
  \chi^2 = \sum_j\left[\frac{\log\Sigma(R_j)-\log\Sigma_{m}(R_j)}{\Delta\log\Sigma(R_j)}\right]^2,
  \label{eq:chi2def}
\end{equation}
with $\Sigma(R_j)$ the observed $\Sigma(R)$ at the $j$-th radial position, $\Delta\log\Sigma(R_j)$ its error, and $\Sigma_{m}(R_j)$ the corresponding model at $R_j$.  The posterior is explored using the   ensemble sampler for Markov Chain Monte Carlo (MCMC) {\tt emcee} \citep[][]{2013PASP..125..306F}. The best fit provided by a least squares routine that minimizes $\chi^2$ was used to initialize the exploration. We carried out a first unconstrained fit,  allowing $f(\epsilon)$ to vary freely. This is used as reference in all the forthcoming discussion. The MCMC exploration was initialized forcing the least squares routine to yield physically sensible solutions with $f(\epsilon)\geq 0$ for all $\epsilon$. Following our previous application of \eimt\, \citep{2024ApJ...973L..15S,2025A&A...694A.283S}, the order  of the polynomial was set to $n=10$, large enough to provide the flexibility needed to reproduce the inner plateau in the stellar distribution observed in galaxies, while resulting in only nine fitting parameters. The  original trial-and-error tests concluded that polynomials of other similar degree yielded equivalent results. The priors in the Bayesian analysis were chosen to be uninformative -- see \citet{2025A&A...694A.283S} for details.  

In essence, \eimt\ returns the function $f(\epsilon)$ required to reproduce the observed $\Sigma(R)$ for a given assumed potential. As it was mentioned above, we consider three potentials: one with a central cusp (NFW) and two with central cores (Schuster–Plummer and $\rho_{230}$). We perform both unconstrained fits, allowing $f$ to take negative values to test for physical inconsistency with the stellar distribution, and constrained fits enforcing $f \ge 0$. The potentials are inconsistent  provided the derivative of the central volume density of stars is close enough to zero (see Sect.~\ref{sec:intro}). Fits forcing $f\geq 0$ enables us to  compare which one of the three potentials best-fits the observed $\Sigma(R)$.

\subsection{Other definitions and conventions employed in the paper}\label{sec:definitions}

The innermost logarithmic slope of the stellar surface density is defined as 
\begin{equation}
  \omega = \lim_{R\to R_i}\frac{d\log\Sigma(R)}{d\log R},
  \label{eq:innermost}
\end{equation}
with $R_i$ the innermost observed radius.  In practice, $\omega$ is computed numerically from the fitted profile interpolated at $R_i$.  

We also use the concept of core radius, here realized as $R_{12}$, defined in terms of $\Sigma(R_i)$ as 
\begin{equation}
  \Sigma(R_{12}) = \Sigma(R_i)/2.
  \label{eq:r12}
\end{equation}

The goodness of the fits assuming different potentials is compared using  the reduced $\chi^2$, $\chi^2/\nu$, where $\chi^2$ is defined in Eq.~(\ref{eq:chi2def}) and $\nu$ stands for number of degrees of freedom of the fit (i.e., the number of observables minus the number of free parameters).

One fundamental aspect of \eimt\ is that a global scaling of the potential does not alter the fit. It changes the scaling of the kernel (Eq.~[\ref{eq:master_mind}]), but this effect is absorbed by a corresponding rescaling of $f(\epsilon)$ (Eq.~[\ref{eq:master}]). A drawback of this property is that the absolute normalization of $f$ remains undetermined, reflecting the lack of constraints on the total mass generating the potential. An advantage, however, is that a global scaling of $\Sigma(R)$ does not influence the fit. Consequently, the results from \eimt\ are insensitive to observational uncertainties such as the distance to the galaxy or the mass-to-light ratio adopted when the mass distribution is inferred from star counts.

Throughout this paper, when we refer to the DM distribution or DM potential, we actually mean the combined contribution of DM and stars. This slight abuse of terminology is not severe, since the low-mass galaxies for which this technique is particularly suited are DM dominated, and the stellar contribution can safely be neglected.

\section{Globular Clusters}\label{sec:globular}
\begin{figure*}
\centering
\includegraphics[width=0.45\linewidth]{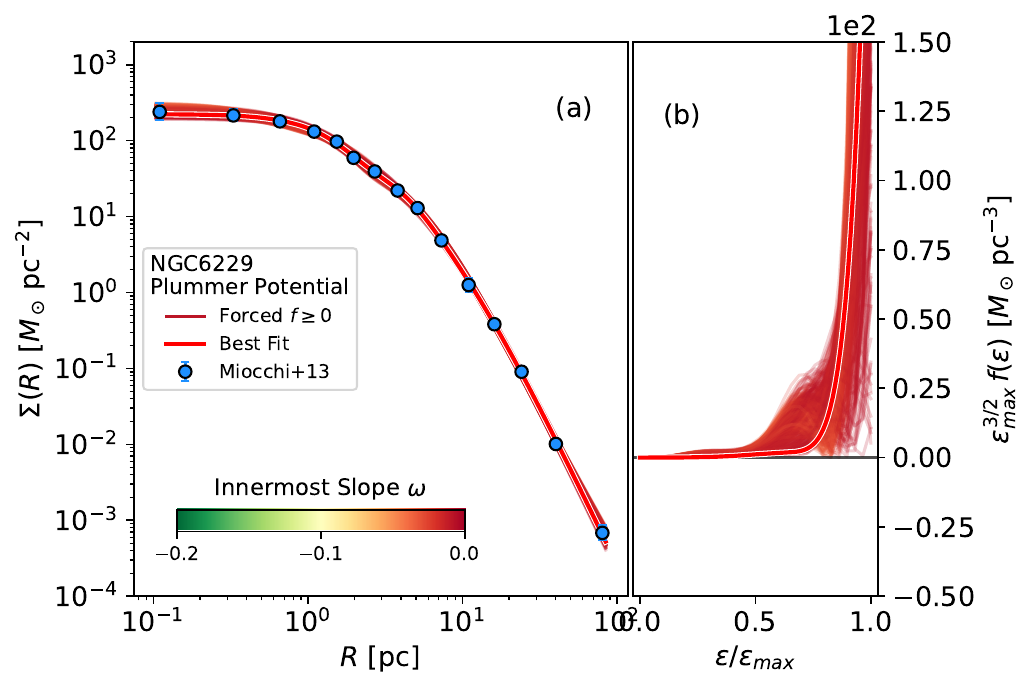}
\includegraphics[width=0.45\linewidth]{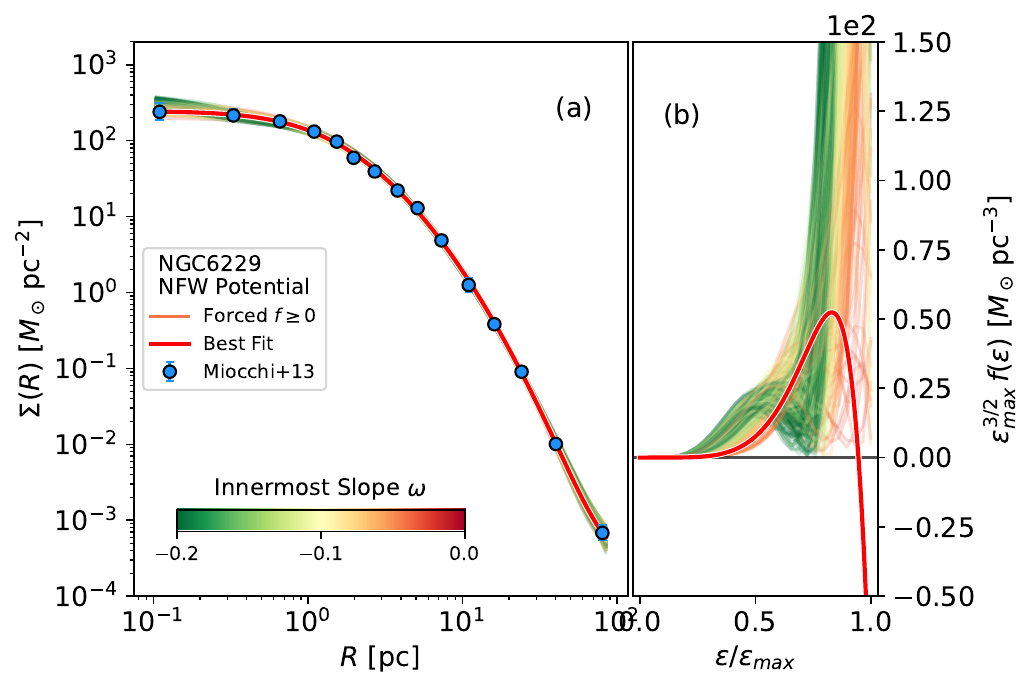}
\includegraphics[width=0.45\linewidth]{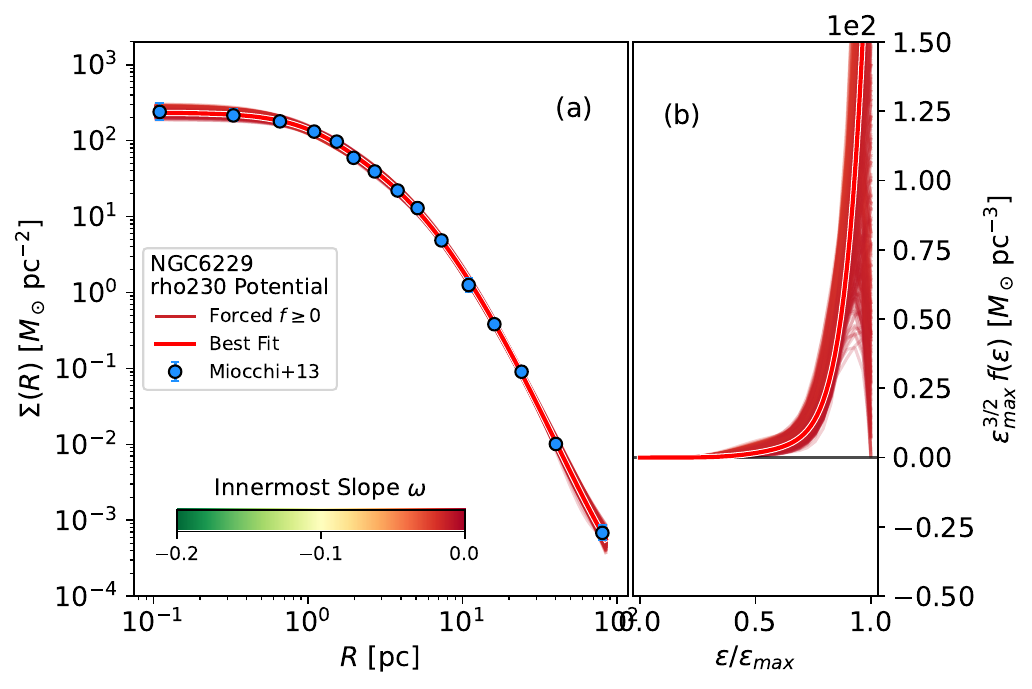}
\includegraphics[width=0.40\linewidth]{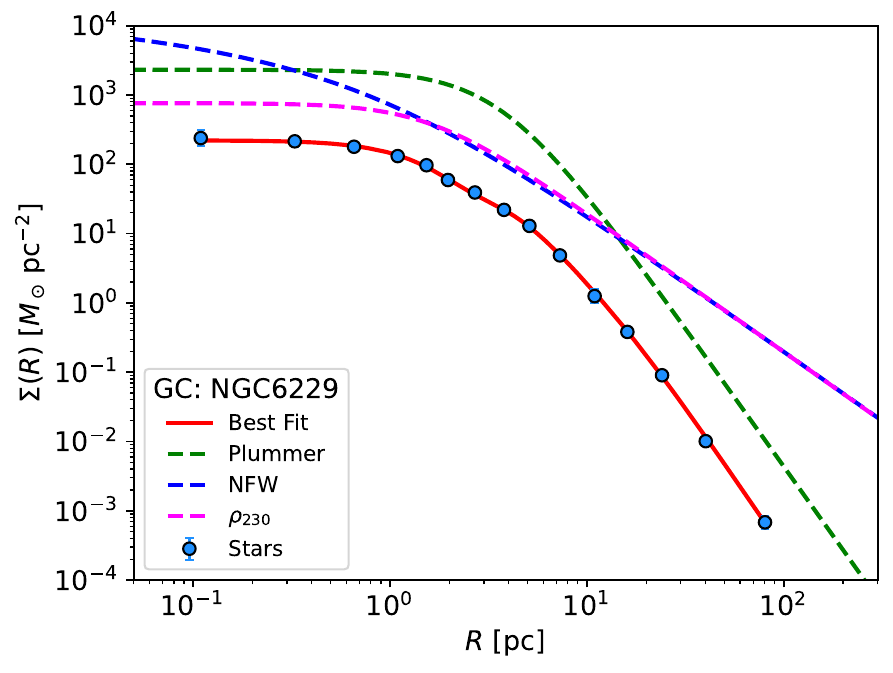}
\caption{
  Example of the typical fit \eimt\ provides for one of the 21 GCs studied in the work (NGC~6229). Top-left panel, label~(a):  fit to the observed stellar mass surface density (blue symbols) assumed to reside in a Schuster-Plummer potential. The red thick solid line represents the least-squares best fit with an unconstrained distribution function $f$,  whereas the other thin lines are the fits derived from the MCMC exploration of the posterior forcing $f$ to be $\geq 0$ $\forall\epsilon$. These other fits are color coded according to the innermost slope of the surface density profile ($\omega$ in Eq.~[\ref{eq:innermost}]; with the equivalence given in the horizontal color bar). Top-left panel, label~(b): distribution functions required by the best fit and by the fits forced to have  $f \geq 0$. The color code is the same as in (a).
Top-right panel, labels (a) and (b): same as the top-left panels, labels (a) and (b), but assuming an NFW potential. Note how the $f$ corresponding to the best fit becomes negative, and so, unphysical. 
Bottom-left panel, labels (a) and (b): same as the top-left panels (a) and (b), but assuming a $\rho_{230}$ potential.
Bottom-right panel: comparison between the stellar surface density profile of the GC and the best-fitting potentials with $f\geq 0$. The plot shows the mass surface density that gives rise to the best-fitting Schuster-Plummer potential (green line), NFW potential (blue line), and $\rho_{230}$ potential (magenta line). The spatial extent of the potentials is set by the fit whereas the vertical scaling is arbitrary, and it was arbitrarily chosen to be 20 times the stellar mass of the GC.
}
\label{fig:df6_run_plot_111}
\end{figure*}
GCs  contain little if any DM \citep[e.g.,][]{2011ApJ...741...72C,2020PASA...37...46B}, therefore, the potential provided by \eimt\ should be consistent with the one produced by the observed distribution of stars. We use the dataset from \citet{2013ApJ...774..151M} providing surface densities for 26 GCs from stellar counts based on HST images. We analyze 21 of them, excluding only those with a few points in the observed radial surface density profiles (AM\,1, Eridanus, Pal\,3, Pal\,4, and Pal\,14). To transform angular separation into physical lengths, we used the distance to the GC listed by \citet{2021MNRAS.505.5957B}. The mass-to-light ratio is assumed to be constant with radial distance. The actual value, set to $1\,M_\odot\,L_\odot^{-1}$,  is irrelevant since the analysis does not depend on a global scaling factor of the stellar surface density profile (Sect.~\ref{sec:definitions}). 

\begin{figure}
\centering
\includegraphics[width=0.85\linewidth]{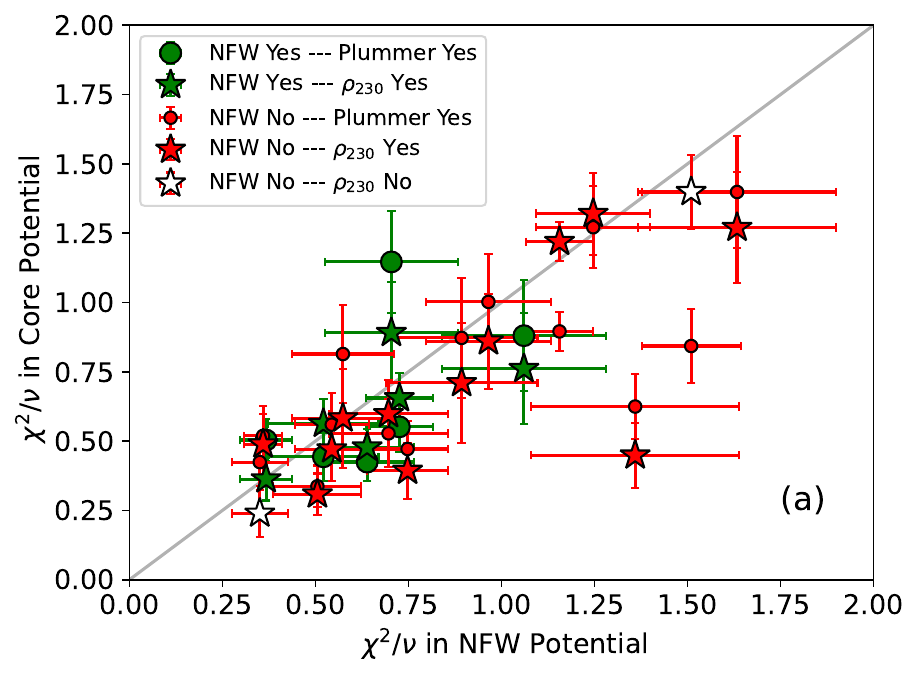}
\includegraphics[width=0.85\linewidth]{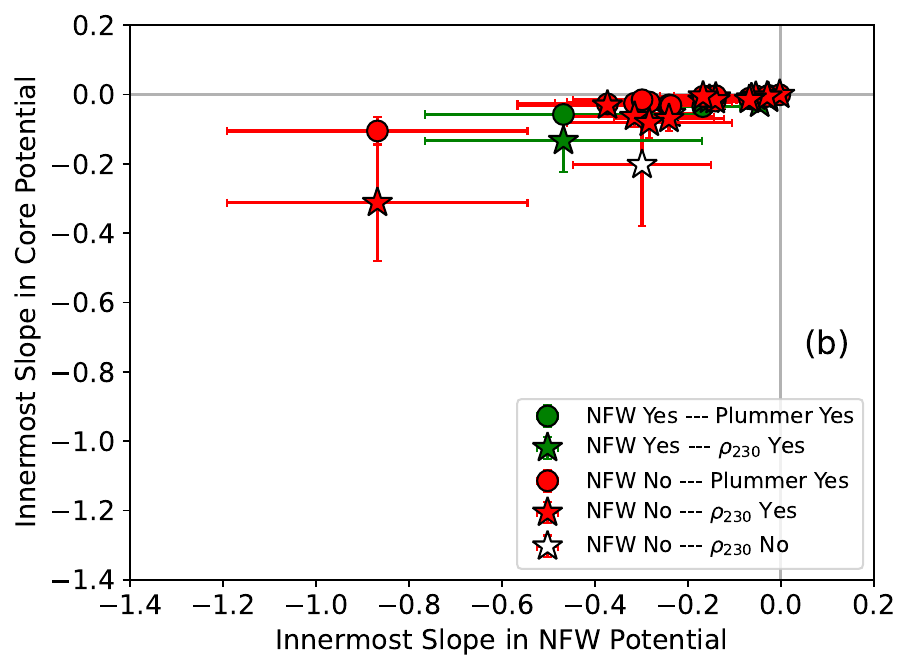}
\includegraphics[width=0.85\linewidth]{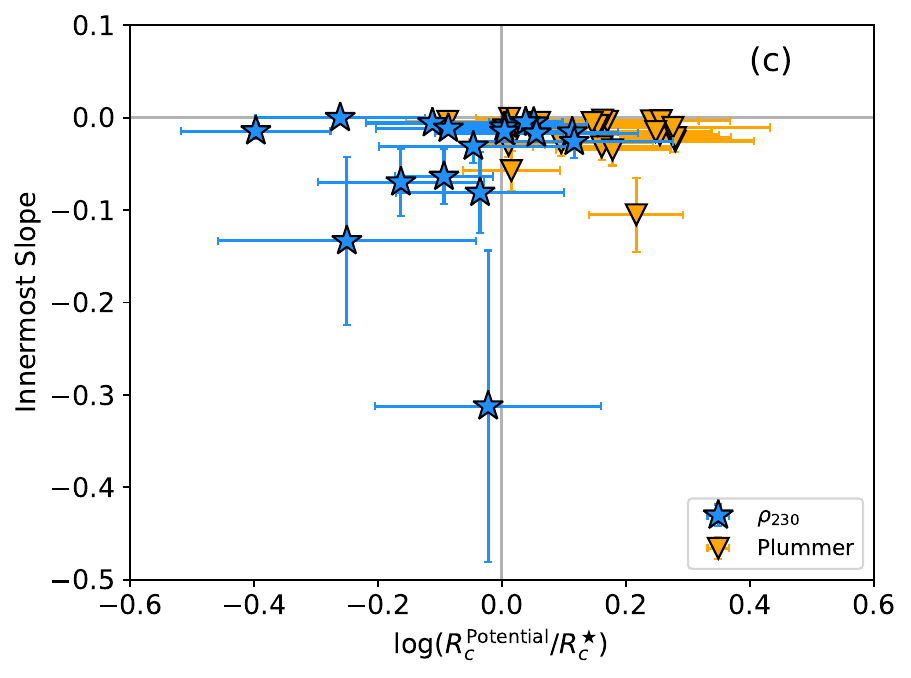}
\caption{
 Summary of the results when \eimt\ is applied to the GCs  observed by \citet{2013ApJ...774..151M}. Each point corresponds to one of the 21 GCs. 
(a) Scatter plot of the reduced $\chi^2$ values obtained with core potentials versus those obtained with NFW potentials, for $f\geq 0$.
 The actual value of $\chi^2/\nu$ depends on the error assumed for the data, which is uncertain, thus, only relative values are significant. What matters is whether the point lies above or below the 1-to-1 relation (gray slanted line).  The color encodes whether for unrestricted $f$, a NFW potential is allowed (green symbols) or forbidden (red symbols). Most symbols are red so most GCs are inconsistent with NFW potentials (71\,\%) whereas all  GCs are consistent with Schuster-Plummer potentials.
 (b) Scatter plot of the innermost slope (Eq.~[\ref{eq:innermost}]) of the fits obtained with core potentials versus NFW potentials. We note how the innermost slope is significantly different from zero in the case of NFW, revealing  an inner upturn not present in the real observation.
 (c)  Scatter plot of the innermost slope versus the ratio between the core radius of the mass accounting for the potential ($R_c^{\rm Potential}$) and core radius for stars ($R_c^\star$). The ratio is reasonably close to one, as expected if the potential is set by the stars.
}
\label{fig:df6_run_plotg}
\end{figure}
Using \eimt ,
we fit the observed surface density of each GCs with the three different gravitational potentials; Plummer-Schuster, NFW, and $\rho_{230}$.  As an example, Fig.~\ref{fig:df6_run_plot_111} gives the fits of one of the GCs (NGC~6229). The top-left panel, label~(a), shows fits to the observed stellar mass surface density (blue symbols) assuming the stars to reside in a Schuster-Plummer potential. The red thick solid line represents the least-squares best fit with an unconstrained $f$,  whereas the other thin lines are the fits derived from the MCMC exploration of the posterior forcing $f\geq 0$ $\forall\epsilon$ (see Sect.~\ref{sec:briefing}). These other fits are color coded according to the innermost slope of the surface density profile ($\omega$ in Eq.~[\ref{eq:innermost}]). The top-left panel, label~(b), represents the distribution function required by  the best fit and by the fits forced to have  $f \geq 0$. The color code is the same as in (a). The  top-right panel, labels (a) and (b), are the same as the top-left panel, labels (a) and (b), but assuming an NFW potential. The  bottom-left panels, labels (a) and (b), are same as before for a $\rho_{230}$ potential. Finally, the bottom-right panel in Fig.~\ref{fig:df6_run_plot_111} compares the stellar surface density profile of the GC with the best-fitting potentials. The plot shows the mass surface density that gives rise to the best-fitting Schuster-Plummer potential (green line), NFW potential (blue line), and $\rho_{230}$ potential (magenta line). The spatial extent of the potentials is set by the fit whereas the vertical scaling is arbitrary and, for the purpose plotting, it was arbitrarily chosen to be 20 times the stellar mass of the GC.

We note that the distribution function $f$ corresponding to the best fit in an NFW potential reaches negative values (Fig.~\ref{fig:df6_run_plot_111}, top-right panel, the red lines), pointing out the inconsistency between the stellar distribution and the cuspy NFW potential. Contrarily, the potentials with cores (Schuster-Plummer and $\rho_{230}$) do not present this problem and are thus consistent with the observed stellar distribution. This behavior represents the most common pattern for the fits to the GCs. Figure~\ref{fig:df6_run_plotg} summarizes the result of the application to all 21 of them. The color in Figs.~\ref{fig:df6_run_plotg}a and \ref{fig:df6_run_plotg}b encodes whether NFW potential
is allowed ($f\ge 0$; green symbols) or forbidden ($f<0$; red symbols) when fits with no restriction in $f$ are carried out.
Most symbols are red so most GCs are inconsistent with NFW potentials (71\,\%) requiring a distribution function partly negative. In contrast, 100\,\%\ of the GCs are consistent with Schuster-Plummer potentials whereas the fraction is 85\,\%\ in the case  of $\rho_{230}$ potentials.
In addition, when $f$ is constrained to be positive, the NFW fits generally produce the largest $\chi^2$ values, indicating that they provide the worst fits among the models considered. In Fig.~\ref{fig:df6_run_plotg}a, the points tend to lie below the 1-to-1 line. NFW fits also tend to develop an inner upturn that is not seen in the observations (see the top-right panel of Fig.~\ref{fig:df6_run_plot_111}).
This upturn is quantified in Fig.~\ref{fig:df6_run_plotg}b by the parameter $\omega$ (Eq.~[\ref{eq:innermost}]), which is found to differ significantly from zero in the case of NFW potentials, but not for Schuster–Plummer or $\rho_{230}$.

Figure \ref{fig:df6_run_plotg}b shows a scatter plot of  $\omega$ versus the ratio between the core radius (Eq.~[\ref{eq:r12}]) of the mass surface density producing the potential and the core radius for stars. The ratio is not exactly one, but reasonably close to this value: $\log(R_c^{\rm Potential}/R_c^\star)=0.15\pm 0.10$ in the case of a Schuster-Plummer potential and $-0.10\pm 0.16$ in the case of $\rho_{230}$ potential (quoted values correspond to the mean and the standard deviation among all GCs).   The small systematic differences between the radii can naturally be attributed to the fact that the stellar distribution is neither a  Schuster-Plummer profile nor a $\rho_{230}$ profile (see, e.g., Fig.~\ref{fig:df6_run_plot_111}, bottom-right panel).   

In short, the application of the \eimt\ to 21 GCs clearly favors the stars to reside in core potentials rather than in cuspy NFW potentials. Moreover, the size of the core of the DM distribution in the core potentials is similar to the observed stellar core size. Thus, the \eimt\ works correctly for GCs.

%
\section{Dwarf spheroidal galaxies with measured potential}\label{sec:dwarfs}

Classical dwarfs have been thoroughly studied over the years, including the use of kinematic data to infer  their DM distribution \citep[e.g.,][]{2009ARA&A..47..371T,2022NatAs...6..659B}. Thus, they also offer a way to test \eimt\ by comparing the constraints it provides with the potentials inferred using full Jeans modeling on these objects. Specifically, our tests targeted three dSphs for which recent studies are available:  Sculptor in Sect.~\ref{sec:sculptor},  Draco in Sect.~\ref{sec:draco}, and Fornax in Sect~\ref{sec:fornax}.   

%
\subsection{Sculptor dwarf galaxy -- data from \citet{2025A&A...699A.347A}}\label{sec:sculptor}

\begin{figure}
\centering
\includegraphics[width=0.9\linewidth]{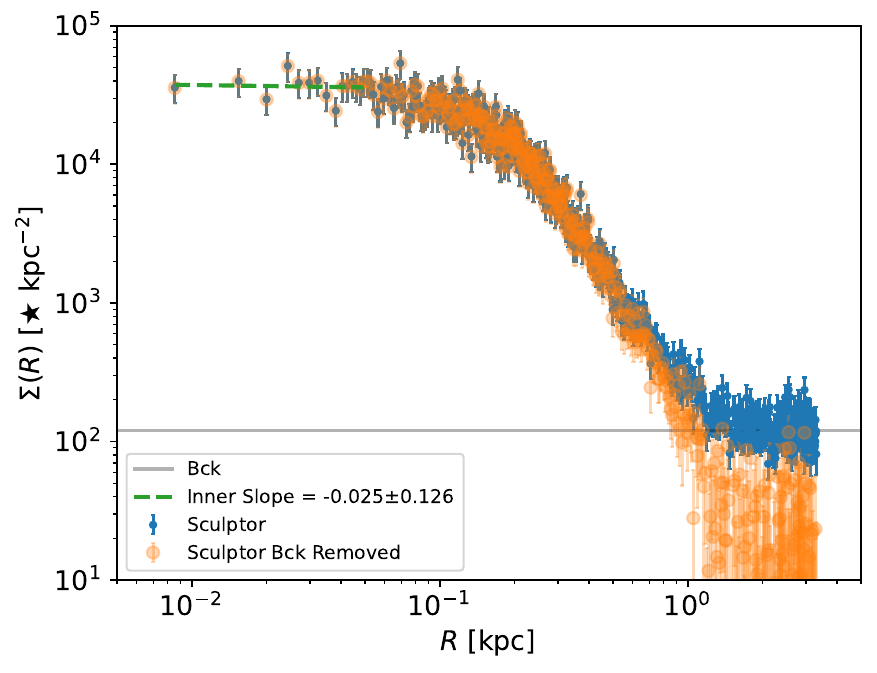}
\caption{
Sculptor surface density profile from \citet{2025A&A...699A.347A}. The background signal was computed from the signal in the largest projected radii (gray solid line), and then subtracted out to go from the original (blue symbols) to the background removed points (orange symbols).  The green dashed line indicates the observed inner slope (Eq.~[\ref{eq:innermost}]), whose value is given in the inset. Projected radii are in kpc and surface densities in stars per kpc$^2$. 
}
\label{fig:read_jmarroyo1}
\end{figure}
The Sculptor dwarf galaxy is a satellite of the Milky Way, located about 90 kpc away and dominated by old, metal-poor stars.
It is dispersion-supported, with little to no gas and strong evidence for a massive DM halo inferred from its stellar velocity dispersion. \citet{2025A&A...699A.347A} carried out a classical Jeans modeling of Sculptor where they infer, among other properties, the DM distribution describing the observed distribution of stars, their position and velocity.
  We construct the observed surface density profile following a procedure similar to, but slightly different from, that in  \citet{2025A&A...699A.347A}. All stars observed by Gaia in the Sculptor region were selected, including foreground contaminants. These contaminants should be subdominant in the central region and, provided they are uniformly distributed across the field, their contribution can be subtracted out.
The selected stars were ordered in radial distance from the center, and then grouped in bins with the same number of objects. The number of stars divided by the area of the ring containing them leads to the surface density profile presented in Fig.~\ref{fig:read_jmarroyo1}. Error bars are assigned assuming Poisson statistics for the star counts, while a constant background is subtracted to remove for the effect of contaminants.

\begin{figure*}
\centering
\includegraphics[width=0.45\linewidth]{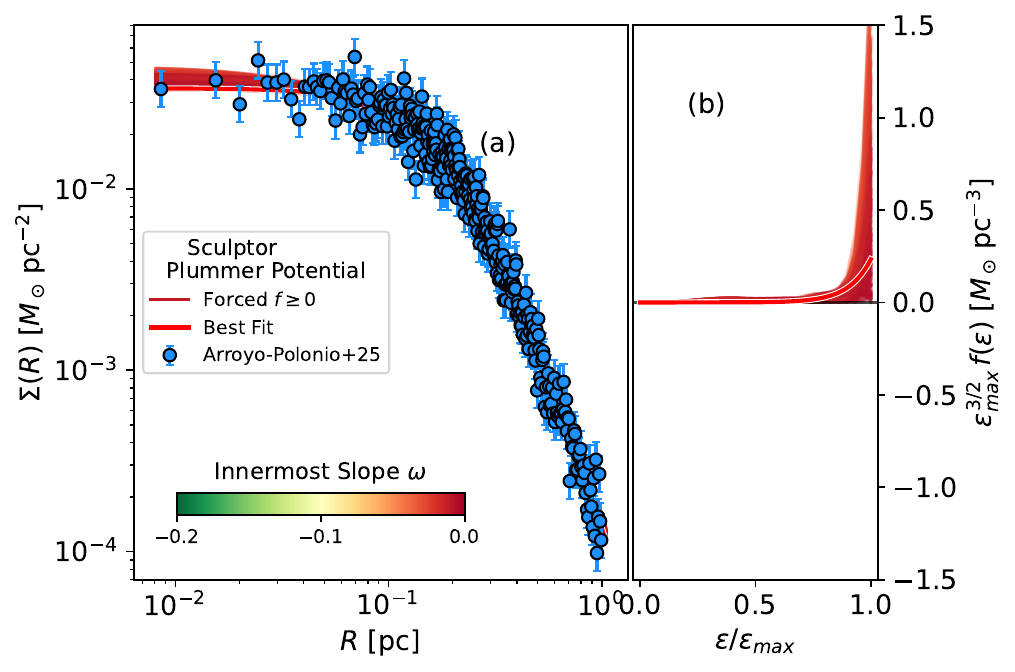}
\includegraphics[width=0.45\linewidth]{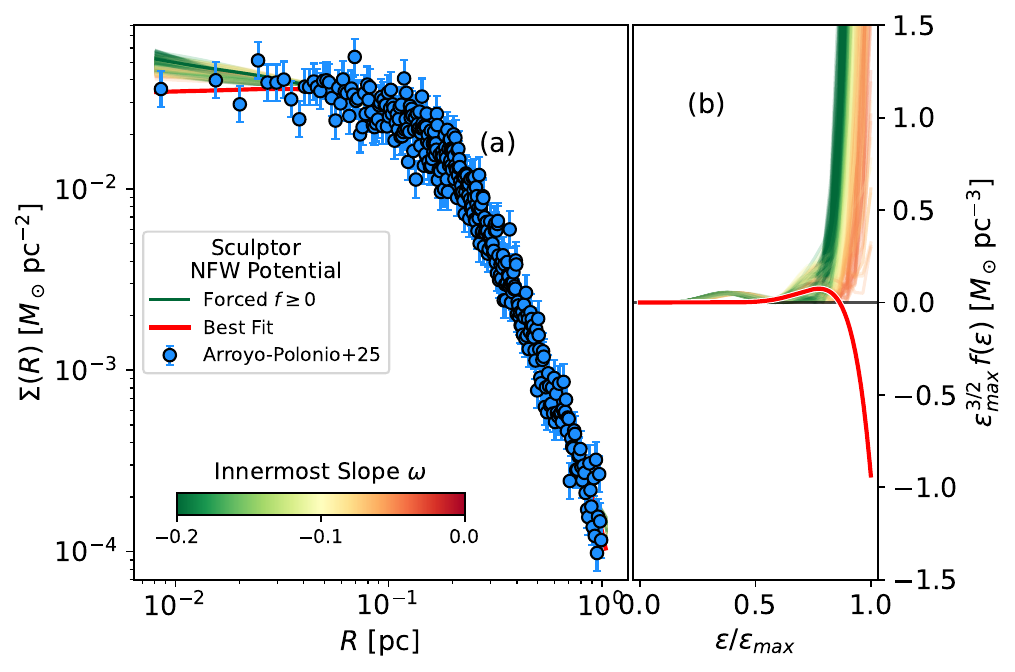}\\
\includegraphics[width=0.45\linewidth]{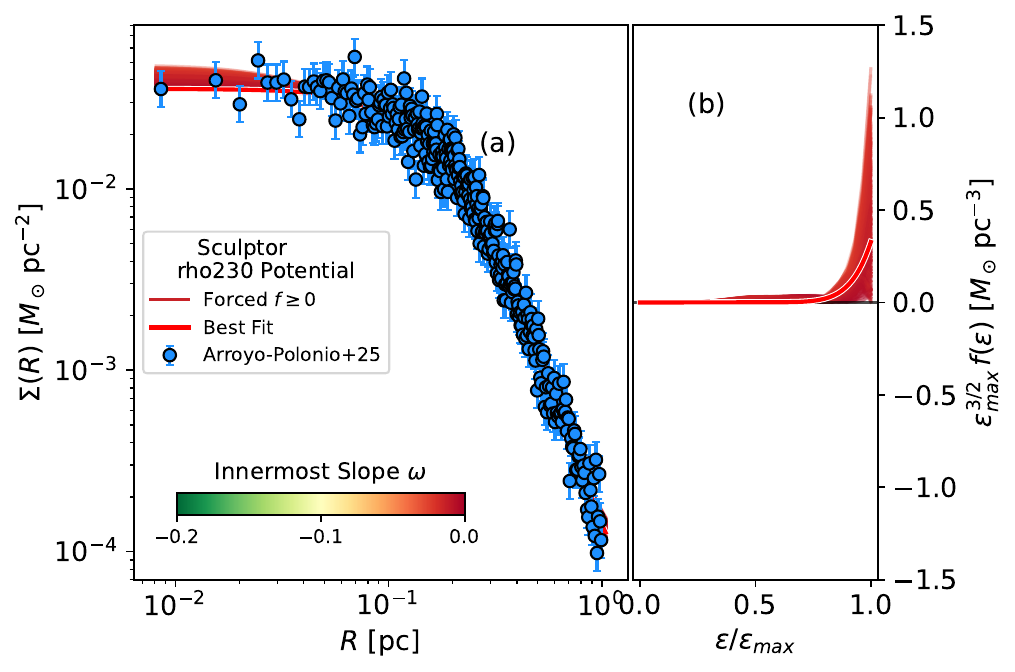}
\includegraphics[width=0.45\linewidth]{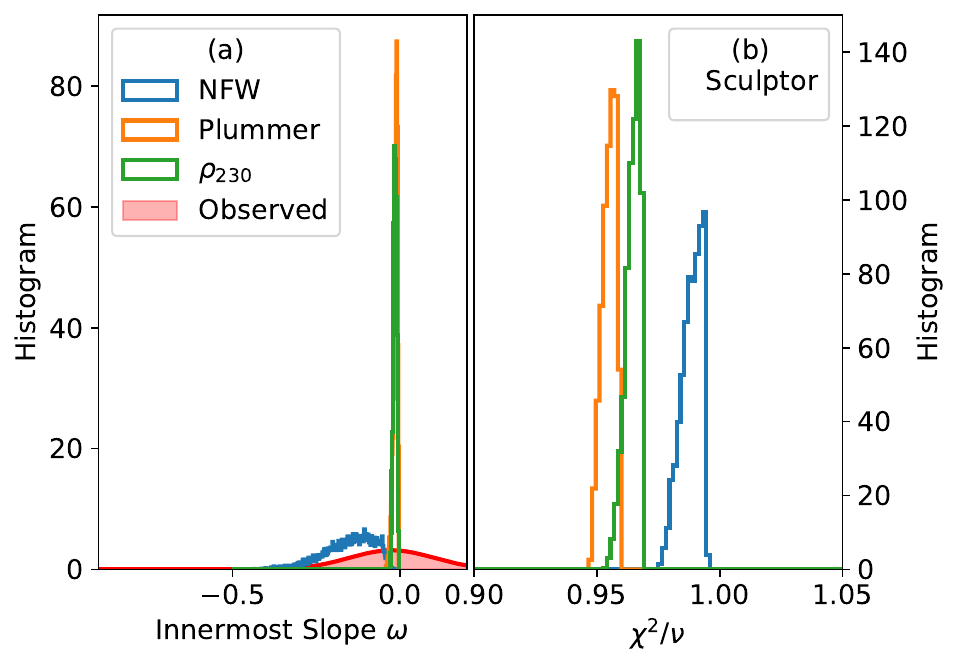}
\caption{
  Application of \eimt\ to the Sculptor data analyzed by \citet{2025A&A...699A.347A} through full Jeans modeling. The layout of the three first panels (top row and bottom-left panel) is the same as those in Fig.~\ref{fig:df6_run_plot_111} and we refer to its caption for details. They show how the unconstrained fit assuming an NFW potential (the solid red line in the top-right panel) prefers $f <0$ and so is unphysical. This problem is not present when assuming Schuster-Plummer or $\rho_{230}$ potentials.   The  bottom-right panel of the figure contains histograms of the reduced $\chi^2/\nu$ and inner slope $\omega$ when the fits are forced to have $f\ge 0$.  
}
\label{fig:df6_run_plot_1_pub}
\end{figure*}
The observed background-removed surface density profile in Fig.~\ref{fig:read_jmarroyo1}  was offered to \eimt , as described in Sect.~\ref{sec:briefing}. The mass-to-light ratio was assumed to be one and independent of the radial position, although the actual value is irrelevant since \eimt\ does not constrain the total DM mass giving rise to the potential (Sect.~\ref{sec:definitions}). The fits resulting from assuming the three potentials are shown in three panels of  Fig.~\ref{fig:df6_run_plot_1_pub}.  The solid red line shows the best fit, leaving $f$ completely unconstrained. The quality of the three fits is similar, reproducing the inner plateau in the stellar distribution at $R \lesssim  0.1 {\rm kpc}$ (Fig.~\ref {fig:read_jmarroyo1}, the dashed green line), but the NFW fit requires an unphysical $f < 0$ (Fig.~\ref{fig:df6_run_plot_1_pub}, top-right panel, the solid red line). If you force the fits to rely on physically meaningful $f \geq 0$ everywhere, then the NFW fit worsens developing an inner slope inconsistent with the observed stellar distribution (the colored thin lines in Fig.~\ref{fig:df6_run_plot_1_pub}, top-right panel). This does not happen with the Schuster-Plummer based fits (Fig.~\ref{fig:df6_run_plot_1_pub}, top-left panel) or $\rho_{230}$ potential (Fig.~\ref{fig:df6_run_plot_1_pub}, bottom-left panel).

The  bottom-right panel of  Fig.~\ref{fig:df6_run_plot_1_pub} summarizes the result of the analysis. It contains histograms of the reduced $\chi^2/\nu$ and inner slope $\omega$ of the fits forced to have $f\ge 0$.  The inner slope of the NFW fits are significantly off the observed value (the dashed green line in Fig.~\ref{fig:read_jmarroyo1}, whose value and error are represented as a red Gaussian in Fig.~\ref{fig:df6_run_plot_1_pub}, bottom-right panel). The inner slopes corresponding to the Schuster-Plummer and $\rho_{230}$ potentials are very close to the observed value.  Moreover,   the reduced $\chi^2/\nu$ of the NFW potential is systematically larger than those for Schuster-Plummer or $\rho_{230}$ potentials. 
\begin{figure}
\centering
\includegraphics[width=0.9\linewidth]{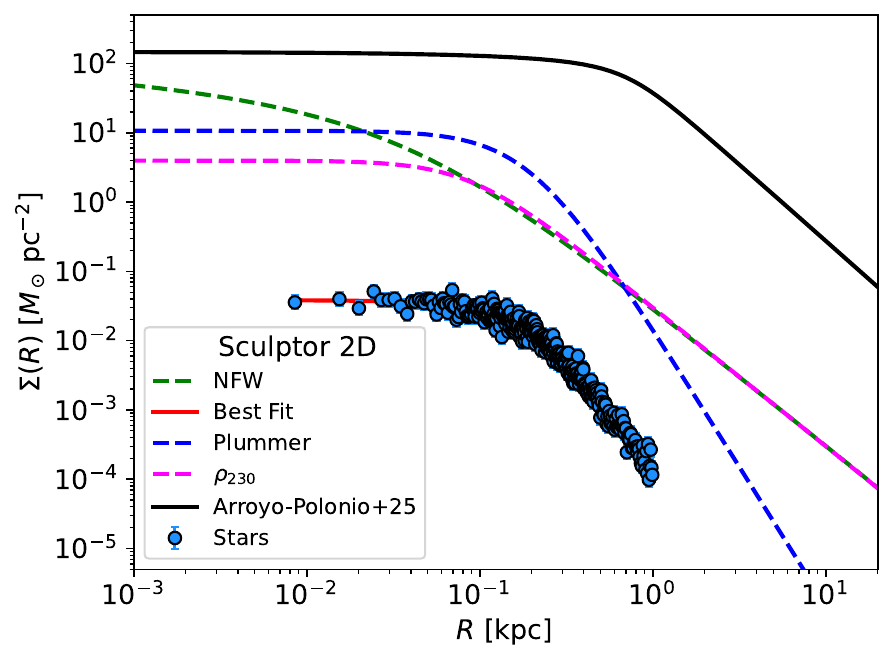}\\
\includegraphics[width=0.9\linewidth]{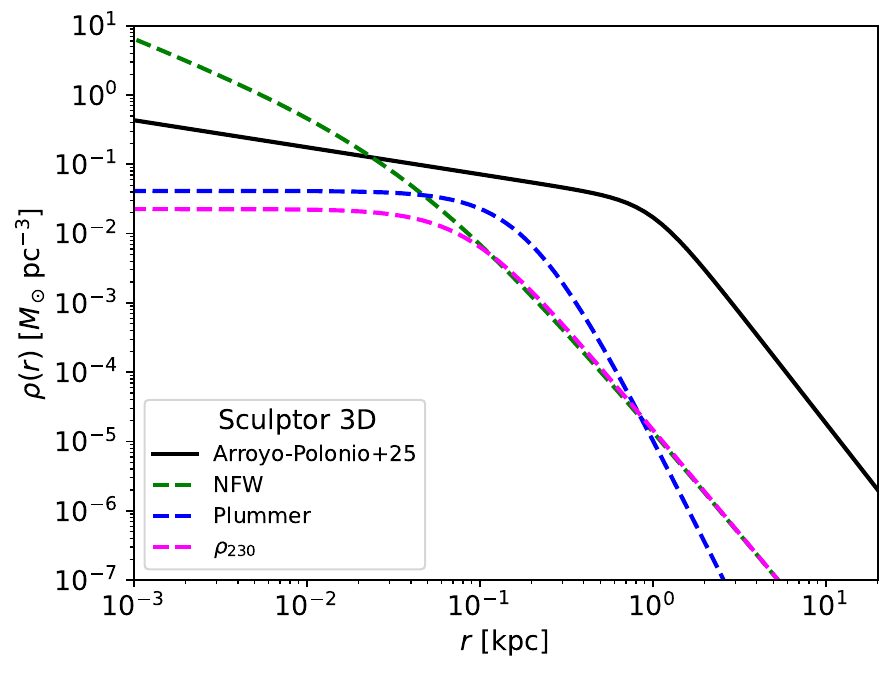}
\caption{
DM halos corresponding to Sculptor. Top panel:  The colored lines show 2D projections of the DM giving rise to the potential that best-fit Sculptor, with the three potentials constrained to have $f \ge 0$. The color code for the potentials is given in the inset. The solid black line represents the 2D projection of the potential that best fit Sculptor according to the analysis by \citet{2025A&A...699A.347A}. The figure also includes the observed stellar distribution (blue symbols).
Bottom panel: 3D distribution of the DM in the three potentials and in the analysis of \citet{2025A&A...699A.347A}.  We note that the overall normalization of the DM distributions remains unconstrained by \eimt, therefore, we have fixed it arbitrarily such that the DM profiles contain 200 times the stellar mass.
}
\label{fig:df6_run_plotd}
\end{figure}
The top panel in Fig.~\ref{fig:df6_run_plotd} shows 2D projections of the DM corresponding to the best fits with the three potentials  constrained to have $f \geq 0$. The solid black line represents the 2D projection of the potential that best-fit Sculptor according to \citet{2025A&A...699A.347A}.  There are three issues to be mentioned:
(1) The DM distribution of both \citet{2025A&A...699A.347A} and the best-fitting Schuster-Plummer profiles do show a core-like inner distribution, very different from the more cuspy NFW profile, (2) the size of the stellar core-like plateau and DM core-like Schuster-Plummer profile are similar, and (3) the size of the core-like feature by \citet{2025A&A...699A.347A} and Schuster-Plummer are clearly different. We get for the Schuster-Plummer 
\begin{equation}
  \log \left(R_c^{\rm Potential}/R_c^\star\right)=-0.08 \pm 0.09,
  \label{eq:gc_radii_ratio}
\end{equation}
whereas in the case of the potential by   \citet{2025A&A...699A.347A}
\begin{equation}
\log \left(R_c^{\rm Potential}/R_c^\star\right)=0.61.
\end{equation}
The bottom panel  in Fig.~\ref{fig:df6_run_plotd} shows the corresponding 3D distributions of DM. The differences between the best-fitting DM NFW profile and the DM from \citet{2025A&A...699A.347A} are even more striking that in 2D.

Thus, \eimt\ indicates that Sculptor is likely not residing in an NFW halo and could be in a cored halo. This conclusion is consistent with the detailed Jeans modeling of the galaxy carried out by \citet{2025A&A...699A.347A}. 


\subsection{Draco dwarf galaxy -- data from Gaia eDR3}\label{sec:draco}

The Draco dwarf galaxy is a dSph satellite of the Milky Way with an heliocentric distance of $\sim\,$76\,kpc. It has a stellar mass $M_\star\sim 5\times 10^5 M_\odot$, a projected half-light radius around 200\,pc,  and it is composed almost entirely of old ($\geq $10 Gyr), metal-poor stars \citep[][]{1998ARA&A..36..435M,2012AJ....144....4M}. Kinematically, Draco exhibits a line-of-sight velocity dispersion of $\sim$10 km\,s$^{-1}$  and little to no measurable rotation, indicating the presence of a strongly DM-dominated halo. 

Draco was chosen because it provides a complementary testbed for \eimt\ since most recent works indicate that its DM halo is cuspy, consistent with NFW potentials \citep{2005MNRAS.363..918L,2013ApJ...763...91J,2018MNRAS.481..860R,2025ApJ...993..249Y,2026Univ...12..195S}. It thus complements the other tests in this paper, where the gravitational potential tends to be sourced by DM distributed in a cored profile.
%

\begin{figure*}
\includegraphics[width=0.45\linewidth]{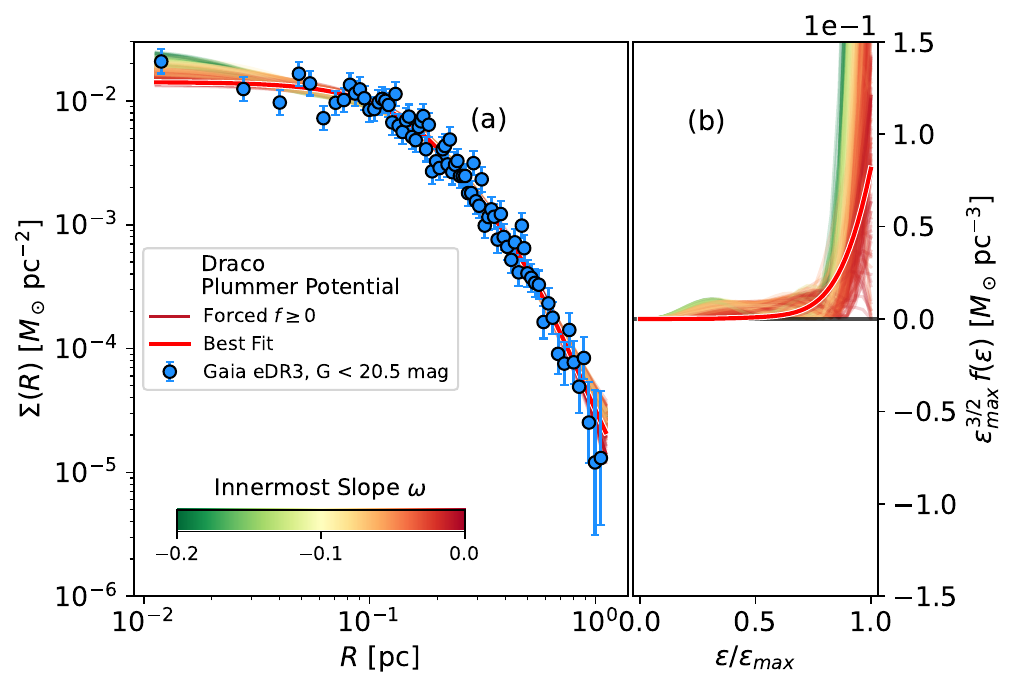}
\includegraphics[width=0.45\linewidth]{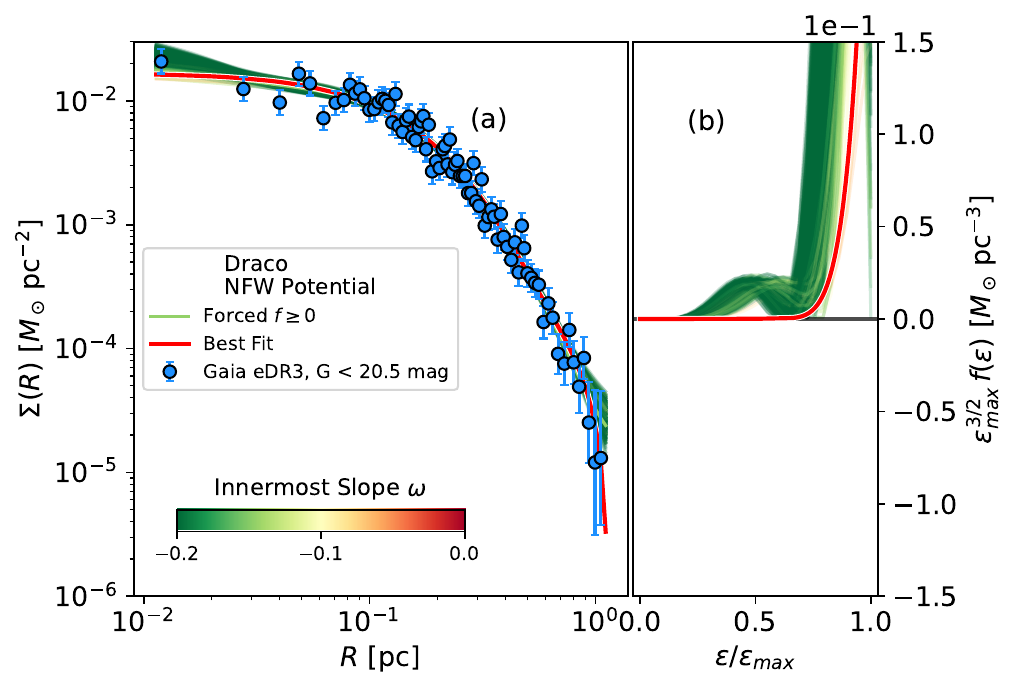}\\
\includegraphics[width=0.45\linewidth]{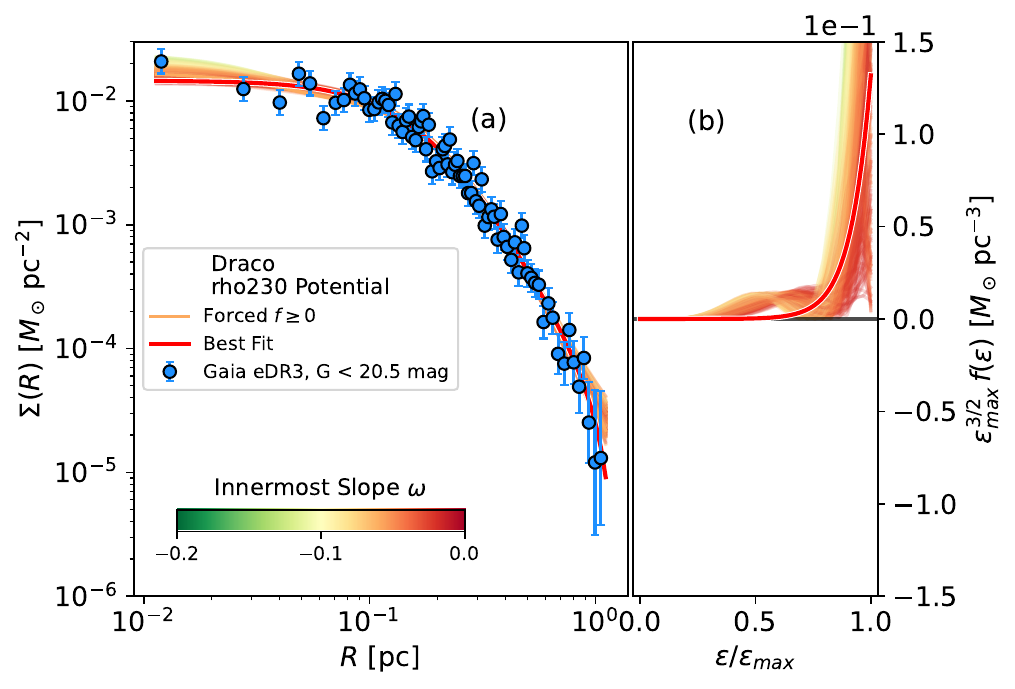}
\includegraphics[width=0.45\linewidth]{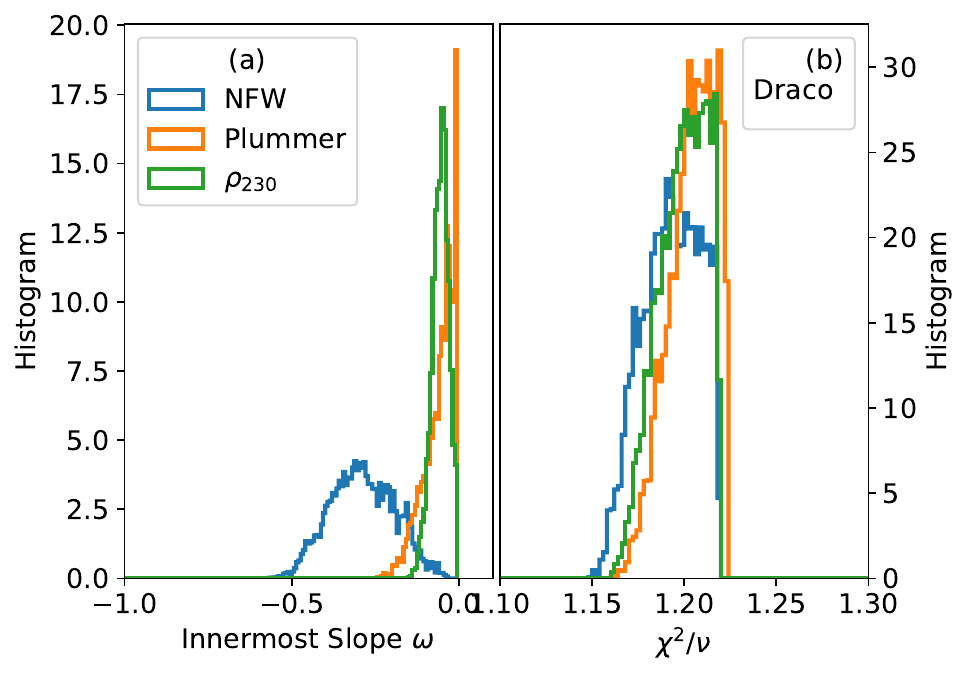}
\caption{
Application of \eimt\ to the stellar surface density of Draco obtained from Gaia eDR3 (see Appendix~\ref{app:draco} for details). The layout of the first three panels (top row and bottom-left panel) is the same as in Fig.~\ref{fig:df6_run_plot_111} and we refer to its caption for details. Here, however, the unconstrained fit assuming an NFW potential (red solid line in the top-right panel) does not prefer $f<0$, and the possibility of the stars residing in an NFW halo is physically realizable. The  bottom-right panel of the figure contains histograms of the reduced $\chi^2/\nu$ and inner slope $\omega$ when the fits are forced to have $f\ge 0$. Note that NFW fits tend to have slightly smaller $\chi^2/\nu$, in stark contrast with the fits to Sculptor (Fig.~\ref{fig:df6_run_plot_1_pub}, bottom-right panel).
}
\label{fig:df6_run_plot_13_pub}
\end{figure*}

The stellar surface density mass of Draco was derived using stellar counts from Gaia eDR3 \citep{2021A&A...649A...1G}, with the selection criteria developed by \citet{2022A&A...657A..54B}. A detailed account of the procedure is given in Appendix~\ref{app:draco}. As in the other tests, we fit the observed mass surface density assuming the three different potentials described in Sect.~\ref{sec:briefing}. The result of these fits are in Fig.~\ref{fig:df6_run_plot_13_pub}. The important difference with respect to the other tests is that the best fit in an NFW potential does not prefer $f$ to be negative -- compare the  top-right panels of  Fig.~\ref{fig:df6_run_plot_13_pub} (Draco; $f\geq  0$) and Fig.~\ref{fig:df6_run_plot_1_pub} (Sculptor; $f < 0$). Thus, the stellar distribution observed in Draco is consistent with the stars being immersed in an NFW DM halo. We note that the observed stellar distribution is also consistent with core potentials (Schuster-Plummer or $\rho_{230}$; Fig.~\ref{fig:df6_run_plot_13_pub}, panels in the left column), which cannot be discarded using \eimt . However, the histogram of reduced $\chi^2/\nu$ (bottom-right panel  of Fig.~\ref{fig:df6_run_plot_13_pub}) seems to favor NFW fits, that present lower values. It has to be compared with the histograms  of $\chi^2/\nu$ for Sculptor in  Fig.~\ref{fig:df6_run_plot_1_pub}, bottom-right panel. Contrarily to what happen with Draco, the largest $\chi^2/\nu$ in Sculptor corresponds to the NFW potential.

\begin{figure}
  \centering
  \includegraphics[width=0.9\linewidth]{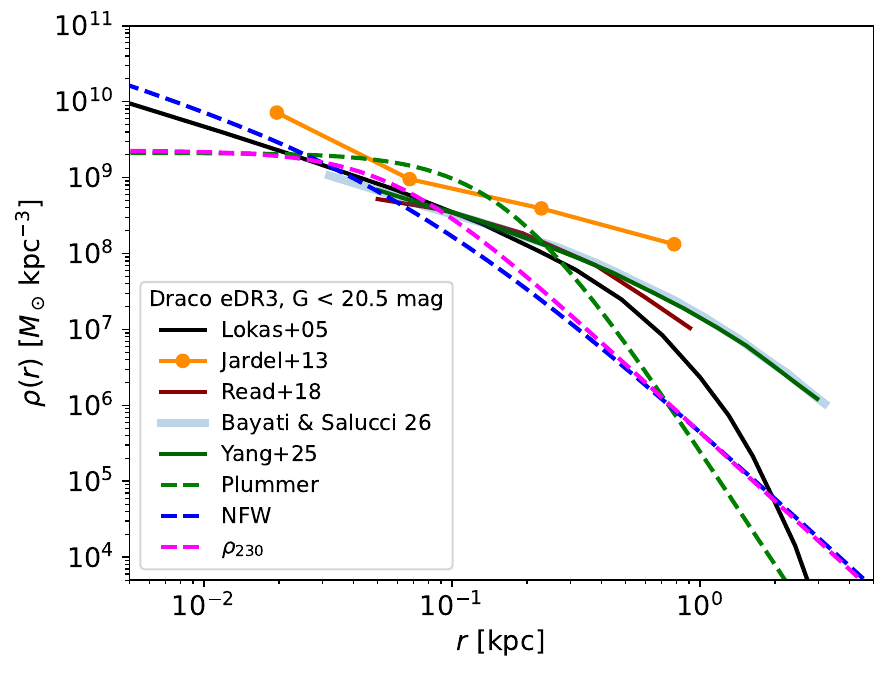}
  \caption{Distribution of DM producing the potential that best-fit the stellar surface density observed in Draco, shown in Fig.~\ref{fig:df6_run_plot_13_pub}; see the dashed lines, with the color given in the inset. According to \eimt , none of them pose a problem of physical incompatibility with the distribution of stars in Draco.
Several other DM distribution found in the literature are also included for reference: Lokas+05 \citep{2005MNRAS.363..918L}, Jardel+13 \citep{2013ApJ...763...91J}, Read+18 \citep{2018MNRAS.481..860R}, Yang+25 \citep{2025ApJ...993..249Y}, and Bayati \& Salucci 26 \citep{2026Univ...12..195S}. All of them show an inner-slope clearly differing from zero.}
\label{fig:df6_run_ploth_battaglia1}
\end{figure}

The fact that \eimt\ shows Draco to be consistent with an NFW potential is in agreement with the works aimed at measuring the DM distribution of Draco, mentioned at the beginning of this section\footnote{Even though \citet{2026Univ...12..195S} question whether the parameters of the NFW halo comply with the predictions of CDM, their best-fitting potential has a NFW shape, which is what \eimt\ examines.}. They tend to show a DM distribution with a cusp rather than a core, namely,  $\rho_{\rm DM}\propto r^{-c}$ with $ 0.5\leq c \le 1$. Four of these external DM profiles are represented in Fig.~\ref{fig:df6_run_ploth_battaglia1} (the inset provides the actual reference). The figure also shows   the DM distribution giving rise to the best fitting potentials for the forced $f > 0$ fits shown in Fig.~\ref{fig:df6_run_plot_13_pub}. According to \eimt , none of them pose a problem to be the gravitational potential holding the distribution of stars in Draco.  

We note that the stellar surface density profiles in Fig.~\ref{fig:df6_run_plot_13_pub} correspond to the selection of Gaia stars with a more lax cut in absolute magnitude ($G < 20.5~ {\rm mag}$). The analysis of the other tighter selection mentioned in Appendix~\ref{app:draco} ($G < 20~{\rm mag}$) is in full agreement with the results presented above.

%
%
\subsection{Fornax dwarf galaxy -- data from \citet{Guerra+26}}\label{sec:fornax}
\begin{figure}
\centering
\includegraphics[width=0.90\linewidth]{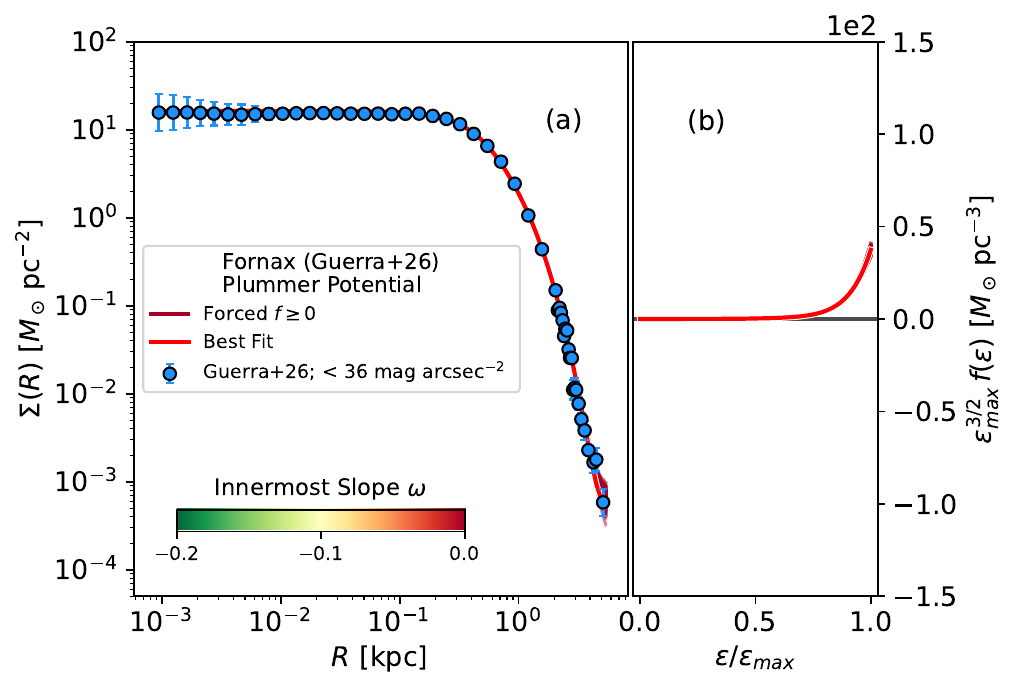}
\includegraphics[width=0.90\linewidth]{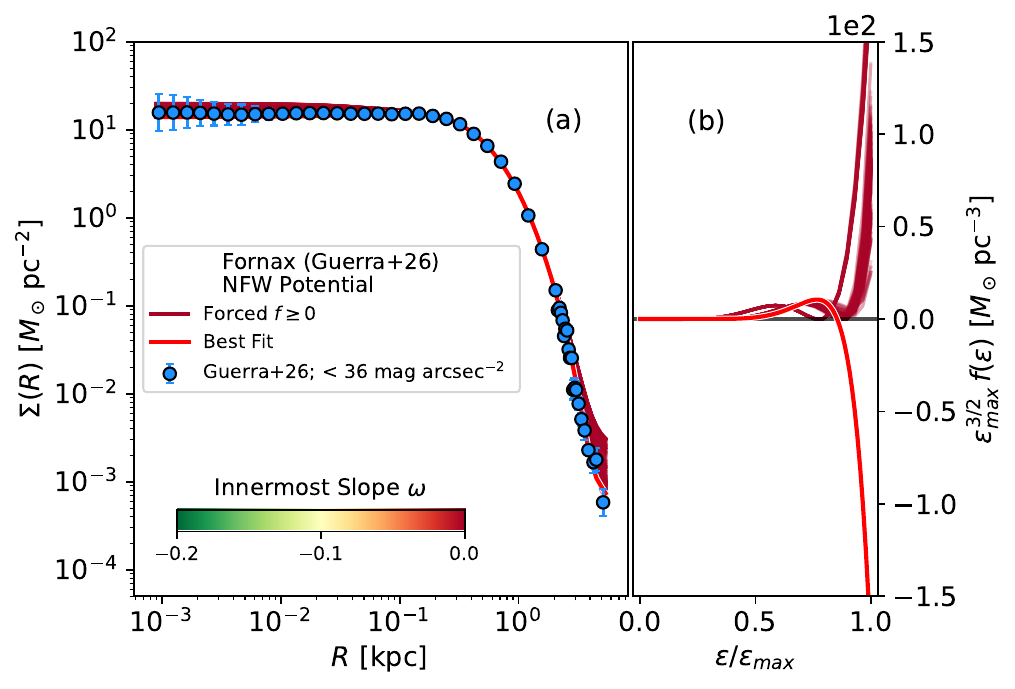}
\caption{
  Application of \eimt\ to Fornax down to  $\mu_V < 36\,{\rm mag\,arcsec}^{-2}$, as observed by \citet{Guerra+26}. Top panels: fit assuming a Schuster-Plummer potential. The best fit is extremely good, with $f$ positive for all $\varepsilon$ (the thick red solid lines). Bottom panels: fit assuming an NFW potential.  In this case the best fit is worst and demands $f<0$, which makes this solution unphysical. The layout is similar to Fig.~\ref{fig:df6_run_plot_111}, and we refer to the caption of this figure for further details. 
}
\label{fig:df6_run_plot_531_pub}
\end{figure}
The Fornax dwarf galaxy is a dSph satellite orbiting the Milky Way  at a distance of about 147\,kpc. It has little gas and no clear spiral structure, but it contains several globular clusters — unusual for a galaxy of its size (half-light radius of around 0.7 kpc). Despite its relatively modest mass ($\sim 3 \times 10^7 M_{\odot}$), it shows evidence of multiple stellar populations, suggesting a more complex evolutionary history than  other dSphs. The extent on the sky is large so that, traditionally, its surface density profile is obtained by star-counting \citep[e.g.,][]{2006A&A...459..423B,2022MNRAS.512.4171Y}.
Recently,  however, \citet{Guerra+26} measured the Fornax surface density profile from spatially integrated photometry using a large field-of-view amateur telescope\footnote{A {\em Stellarvue} telescope with an aperture of 130\,mm and a field-of-view of  $1.5 \times  1.1 \deg^2$, corresponding to a plate scale of 
1.2~\arcsec\,pixel$^{-1}$.}, enough to target all the full galaxy simultaneously, including the innermost regions. They found agreement with the surface brightness derived from star-counting in the outskirts but significant differences in the inner core where crowding could be an issue for star counting. Their exquisite surface brightness profile shows an extended almost flat inner plateau, which we have used to test \eimt .   

\citet{Guerra+26} measured the surface brightness in the  $V$-band, $\mu_V$, down to around 30~mag\,arcsec$^{-2}$, which roughly happens at a distance of 50\,arcmin from the center of the galaxy. The profile was completed down to around 36~mag\,arcsec$^{-2}$ matching their observation with the curve from star-counting by \citet{2022MNRAS.512.4171Y}. For the analysis presented here, the  surface brightness profile was transformed to mass surface density assuming a constant mass-to-light ratio and a stellar mass of  $\log(M_\star/M_\odot)= 7.39$ \citep{2021MNRAS.501.2332B}. For a solar $V$-band absolute magnitude of $4.83$,  this transformation implies a sensible mass-to-light ratio of $\simeq  1.9 M_\odot/L_\odot$. The error bars provided by \citet{Guerra+26} account for the uncertainties in the determination of the center of the galaxy.

The application of the tool with the NFW potential is shown in the bottom panels of Fig.~\ref{fig:df6_run_plot_531_pub}.   The best fit leaving $f$ unconstrained yields $f<0$, therefore, it is unphysical (the red solid lines in the figure). However, when assuming the stars to reside in a Schuster-Plummer potential, then the problem clears out (bottom panels of Fig.~\ref{fig:df6_run_plot_531_pub}). The same happens when  $\rho_{230}$ potentials are used (not shown). In all three cases $\omega\simeq 1$, reproducing quite well this feature of the observed stellar distribution, as it should be if the observed plane-of-the-sky projected stellar distribution is close enough to the center of the distribution \citep{2025arXiv251215886V,SanchezAlmeida26}.

\begin{figure}
\centering
\includegraphics[width=0.8\columnwidth]{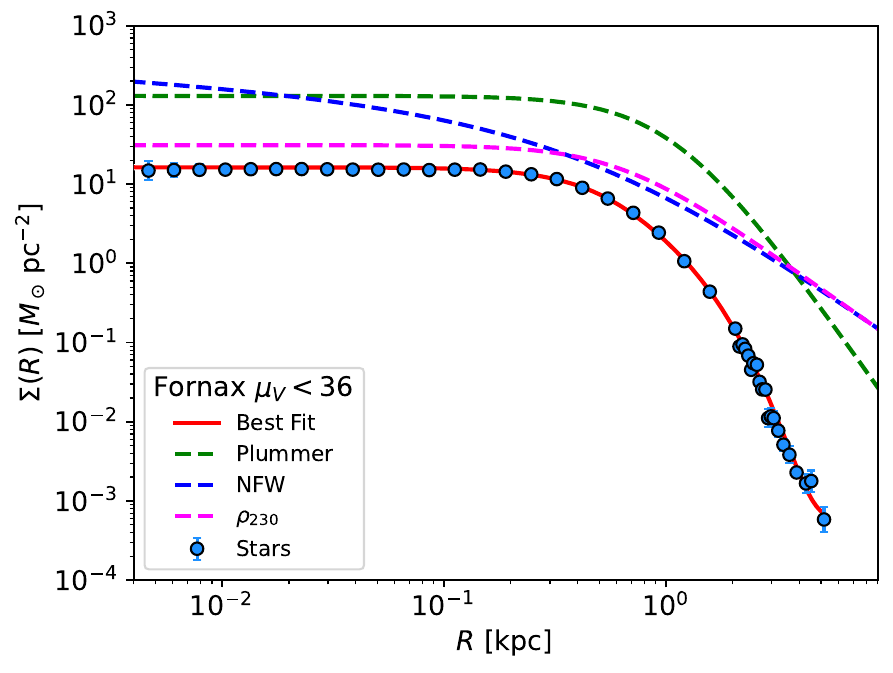}\\
\includegraphics[width=0.8\columnwidth]{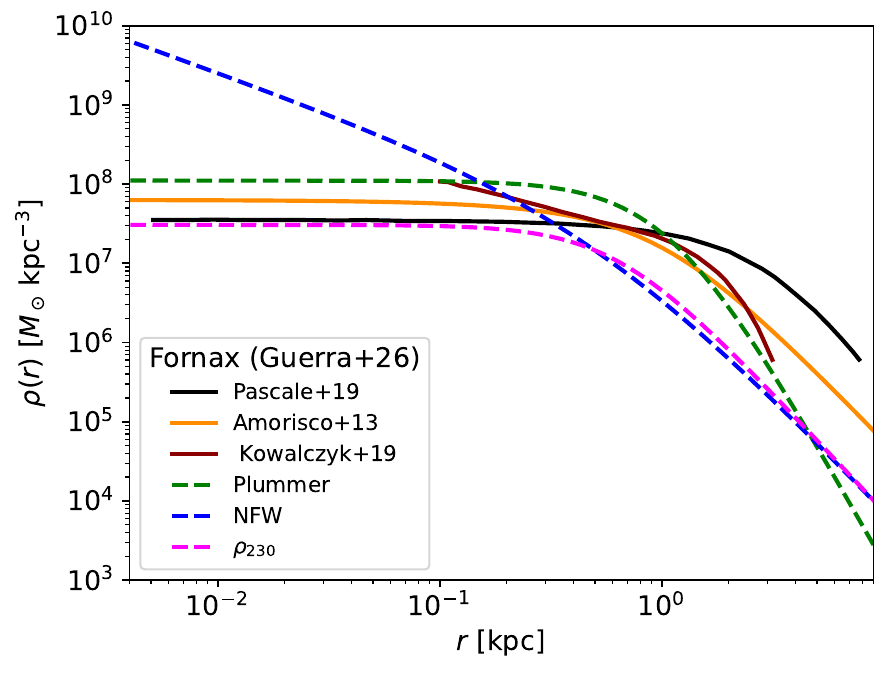}
\caption{
DM distributions consistent with the stellar surface of Fornax, according to \eimt . 
Top panel:  best fit (red line following symbols) together with the DM distribution corresponding to this best fit in the three assumed potentials. The Schuster-Plummer DM profile follows quite well the stellar distribution.  We note that the mass assigned to the potentials is arbitrary, and in this case it was assumed to be 50 times the value of $M_\star$.
  Bottom panel: 3D density corresponding to the DM distributions from the top panel and using the same  color code used in there. The plot also includes three independent recent estimates of the DM distribution in Fornax based on stellar kinematics. 
}
\label{fig:df6_run_plotd_fornax_sguerra3}
\end{figure}
The top panel in Fig.~\ref{fig:df6_run_plotd_fornax_sguerra3} shows the observed stellar distribution together with the 2D projection of the DM distribution corresponding to the best fit in the three assumed potentials. Note that the Schuster-Plummer DM mass profile follows quite well the observed stellar distribution.
The bottom panel of  Fig.~\ref{fig:df6_run_plotd_fornax_sguerra3} shows the 3D density corresponding to these best-fitting DM distributions (the colored dashed lines, with the same color code as in the upper panel). This bottom panel also includes three independent recent estimates of the DM distribution in Fornax based on Jeans analyses that considers stellar kinematics. The ones by \citet{2013MNRAS.429L..89A} and \citet{2019MNRAS.488.2423P} show an inner core (the black and orange solid lines), in qualitative agreement with the Schuster-Plummer potential that best fits the stellar distribution.  The DM distribution inferred by \citet{2019MNRAS.482.5241K}, based on the  Schwarzschild orbit superposition method, does not show the core so clearly (the dark red line), but it is definitively not an NFW profile. Moreover, the difference could be due the data they use. \citeauthor{2019MNRAS.482.5241K} used a dataset which does not show the conspicuous extended core of Fornax present in the data employed in our test.

%
%
\begin{figure*}
\centering
\includegraphics[width=0.45\linewidth]{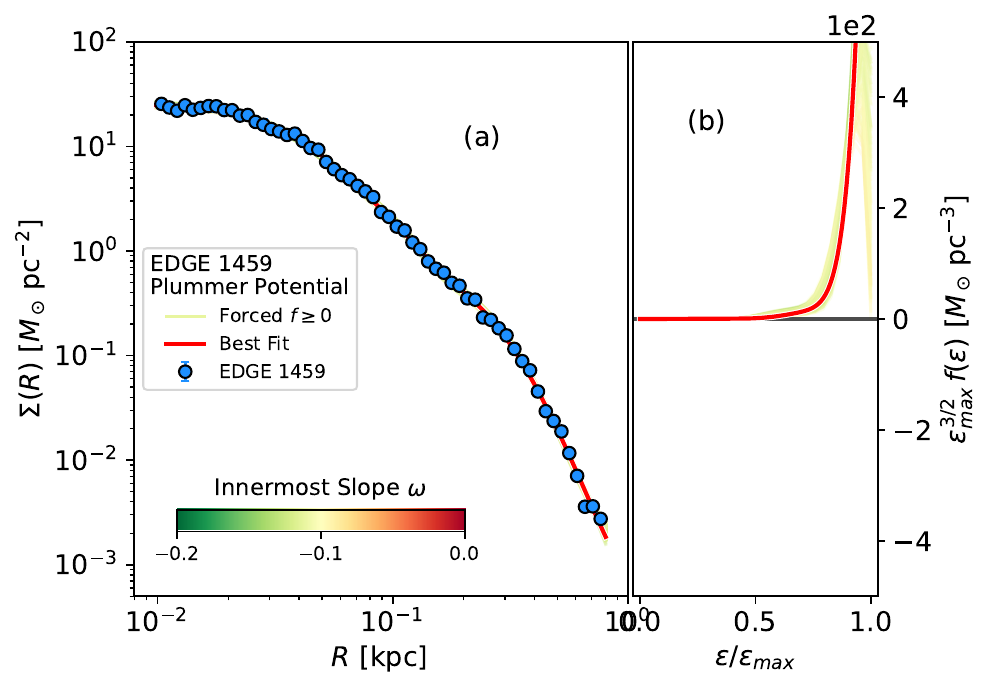}
\includegraphics[width=0.45\linewidth]{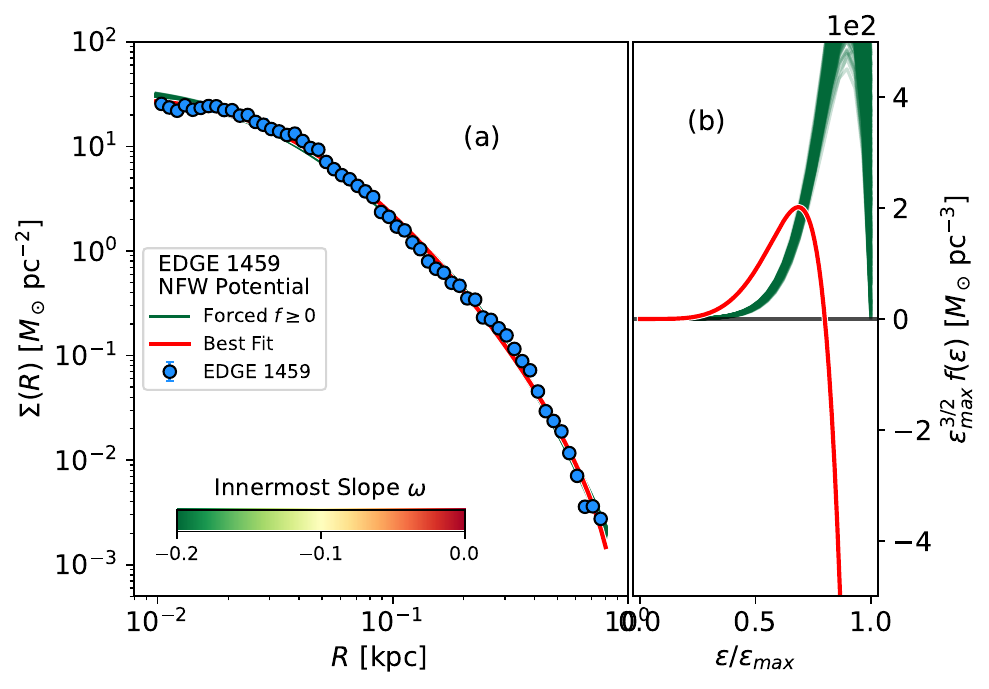}\\
\includegraphics[width=0.45\linewidth]{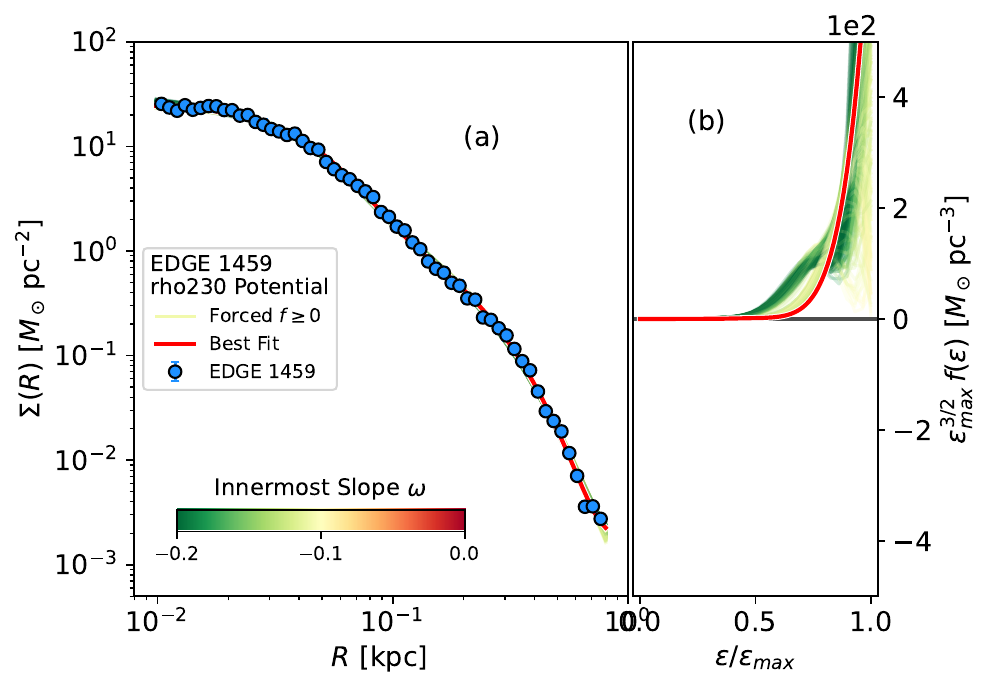}
\includegraphics[width=0.40\linewidth]{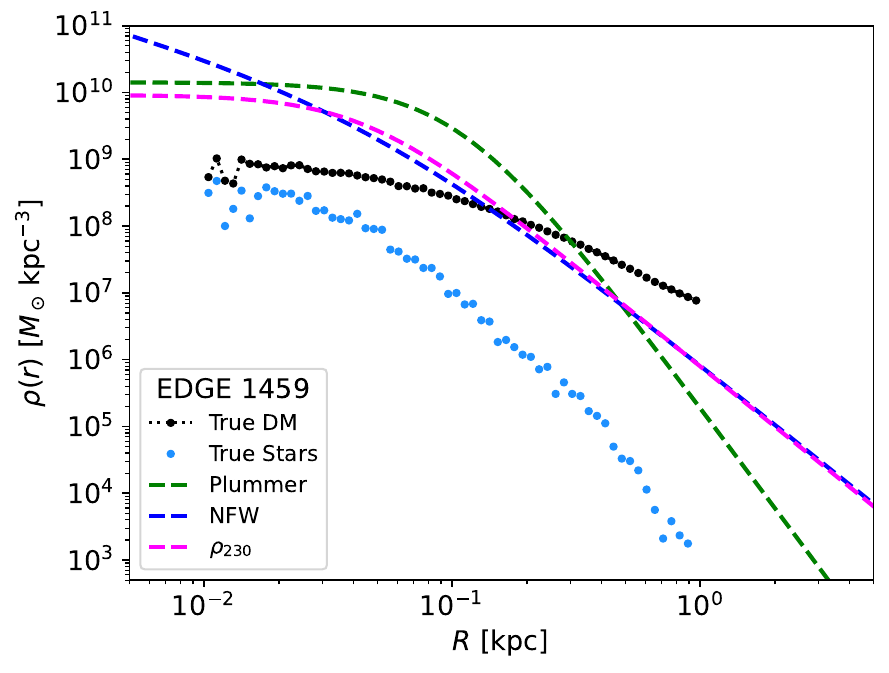}
\caption{
Fits corresponding to the surface density of a simulated galaxy of known DM distribution (specifically, EDGE~halo~\#1459). Top-left panel: fit (label a) and $f$ (label b) when the potential is assumed to be of Schuster-Plummer shape.  The layout is similar to that in Fig.~\ref{fig:df6_run_plot_111}, and we refer to its caption for details. Top-right panel: same as the top-left panel but assuming the potential to be NFW whose best fit requires $f <0$ (the solid red lines).  Bottom-left panel: same as the panels in the top row but assuming the potential to be of the type $\rho_{230}$. Bottom-right panel: True three-dimensional DM distribution (black symbols), which should be compared with the best-fitting distributions corresponding to the three different potentials for $f \ge 0$ (the dashed lines, as labelled in the inset). We note that the overall normalization of the DM distribution is unconstrained by \eimt, therefore, we have fixed it arbitrarily such that the profiles contain 200 times the stellar mass.
}
\label{fig:df6_run_plot_411_pub}
\end{figure*}
\section{Application to numerical simulations where the true DM distribution is known}\label{sec:edge}

The DM distribution in numerical simulations is known. Therefore, galaxies from simulations provide an ideal testbed for \eimt, since the constraints it provides can be directly compared with the true underlaying DM distribution. We have chosen the EDGE suite \citep[Engineering Dwarfs at Galaxy Formation’s Edge; ][]{2020MNRAS.491.1656A,2025MNRAS.541.1195R} for this purpose, as it aims to realistically model the smallest DM-dominated dwarf galaxies, which are good proxies for the type of object targeted by \eimt.
A complementary test based on Feedback In Realistic Environments simulations \citep[FIRE;][]{2018MNRAS.480..800H}   is analyzed in Appendix~\ref{app:fire}, rendering results fully consistent with those presented  here.

The EDGE simulations are a suite of high-resolution cosmological zoom-in simulations of dwarf galaxies at the threshold of galaxy formation, spanning DM halo masses between $10^9$ and $10^{10}\,M_\odot$. They are run with a radiation-hydrodynamics code that includes gas cooling, star formation, and stellar feedback (supernovae, radiation pressure, and photoionization), as well as a treatment of the cosmic UV background. The simulations achieve pc-scale spatial resolution and particle masses of the order of $10^2\,M_\odot$, allowing them to resolve the multiphase inter-stellar medium and bursty star formation in ultra-faint systems. 

We selected two DM halos with enough stars to provide well-defined stellar surface density profiles ($M_\star\sim 10^5\,M_\odot$) also having large DM mass within the stellar distribution ($\sim 500$ times $M_\star$). The volume  density in DM and stars was computed counting particles in shells around the center. These halos are representative of the low stellar mass halos in EDGE \citep{2021MNRAS.504.3509O}. 
The fits for one of them is shown as an example in Fig.~\ref{fig:df6_run_plot_411_pub}. We assume a constant error to carry out the fit, adjusted so the final reduced $\chi^2/\nu$ is approximately one. As one can see,  cored potentials are compatible with the observed stellar distribution (Fig.~\ref{fig:df6_run_plot_411_pub}, left column panels) but  the best fit in NFW potentials is not (Fig.~\ref{fig:df6_run_plot_411_pub}, top-right panel -- the red line has $f < 0$). Moreover, when $f$ is forced to be positive, the NFW-based fits worsen considerably compared with the fits assuming  cored potentials:  $\chi^2/\nu= 1.7\pm 0.1$ (NFW), $0.6\pm  0.1$ (Schuster-Plummer), and  $0.8 \pm 0.1$ ($\rho_{230}$). The inner slope of the NFW fit is also larger than observed and larger than those from cored potentials: $\omega =  -0.33\pm 0.05$ (NFW), $-0.11\pm 0.03$ (Schuster-Plummer), and $-0.14\pm 0.13$ ($\rho_{230}$).
Based on these results from \eimt, we conclude that the DM halo hosting the stars is not well described by an NFW potential. We also find that the inferred surface density is consistent with cored potentials. The bottom-right panel of Fig.~\ref{fig:df6_run_plot_411_pub} shows the true three-dimensional DM distribution (black symbols), which can be compared with the best-fitting distributions corresponding to the three different potentials for $f \ge 0$ (dashed lines, labeled in the inset). As predicted by \eimt, the true DM distribution deviates significantly from the NFW profile that best fits the stellar distribution. If anything, the true potential exhibits a core, although it is more extended than those of the best-fitting models, and also more extended than the stellar distribution itself (blue dots in Fig.~\ref{fig:df6_run_plot_411_pub}, bottom-right panel).

We note that the conclusions drawn from fitting the second EDGES halo are similar to those described above, and are also consistent with those derived for the FIRE low-mass halo analyzed in Appendix~\ref{app:fire}.

%
%
\section{Discussion and conclusions}\label{sec:discussion}

Galaxies with stellar mass below a $10^{5-6} M_\odot$ threshold are expected to retain their primordial cusps under CDM, which makes the analysis of these low-mass systems a particularly powerful diagnostics. If DM cores are found in such systems, it would likely signal genuine departures from CDM. \citet{2025A&A...694A.283S} introduce a diagnostic tool (\eimt ) based on the Eddington Inversion Technique which rather than relying on expensive multi-epoch spectroscopy, uses only photometry. The key idea is that if stars with a cored distribution are placed in a cuspy NFW potential, the phase-space distribution function $f$ required to sustain that configuration becomes negative -- which is physically impossible. This inconsistency is what \eimt\ exploits. The observed stellar surface density is fitted using $f$ as free parameter once a particular gravitational potential is assumed (Sect.~\ref{sec:briefing}). Three different gravitational potentials are considered: an NFW potential, a Schuster–Plummer potential, and a $\rho_{230}$ potential. When the best-fit choses $f < 0$, then \eimt\ tells that the assumed potential is inconsistent with the observation. When $f$ in \eimt\ is forced to be positive, $\chi^2$ and properties like the inner slope of the fit (Eq.~[\ref{eq:innermost}]) indicate preference of one potential with respect to the others. 

Following the introduction of \eimt\ in Paper~I \citep{2025A&A...694A.283S}, this second paper in the series presents a set of tests of the method using observed and simulated systems with independently known gravitational potentials. The goal is to compare the constraints inferred by \eimt\ with potentials established through other means: systems that are effectively self-gravitating, galaxies previously analyzed with classical dynamical techniques, and numerical simulations in which the DM distribution is computed self-consistently. The full set of tests considered in this work is listed in Table~\ref{table:summary}.

\eimt\ is applied to a sample of 21 GCs in Sect.~\ref{sec:globular}, which are known to have self-gravitating stellar distributions and central plateaus in their stellar surface density  (see Fig.~\ref{fig:df6_run_plot_111}).  The study reveals that the NFW model is physically incompatible with 71\,\% of the sampled clusters requiring an unphysical negative $f$ to explain the data. In stark contrast, 100\,\% of the clusters were found to be consistent with the cored Schuster-Plummer potential. The analysis further shows that when $f$ is restricted to physically valid positive values, the NFW fits fail to reproduce the observed stellar profiles, instead creating an artificial inner upturn that is not present in the real observations. Additionally, \eimt\ accurately retrieved potential core radii that were consistent with the observed stellar core radii, yielding a ratio close to unity (Eq.~[\ref{eq:gc_radii_ratio}]) as expected for systems where stars determine the gravitational well. Consequently,  these findings validate the ability of \eimt\ to correctly identify the cored nature of gravitational potentials in GCs using only photometry.

Section~\ref{sec:dwarfs}  applies \eimt\ to three dSphs, namely, Sculptor, Draco, and Fornax, to validate the technique against gravitational potentials independently determined through traditional kinematic analyses and Jeans modeling.  In the case of Sculptor (Sect.~\ref{sec:sculptor}), the tool reveals that an NFW potential is physically incompatible with the observed stellar distribution, requiring either $f < 0$ or, when restricted to positive values, producing an inner slope that fails to match the observed central plateau. These findings favor a cored halo, which is in direct agreement with independent detailed Jeans modeling. Conversely, Draco (Sect.~\ref{sec:draco}) provides a critical test case for cuspy systems; \eimt\ finds its surface density consistent with an NFW potential, matching previous results in the literature and demonstrating the ability to signaling cusps. Finally, the analysis of Fornax (Sect.~\ref{sec:fornax}) shows that its extended inner stellar plateau is incompatible with an NFW profile, further corroborating existing evidence for a DM core in this system.  Overall, these results demonstrate that \eimt\, using only photometric data, consistently reproduces the conclusions of more precise kinematic studies for dSph galaxies.

Section~\ref{sec:edge} evaluates \eimt\ using numerical simulations of EDGE galaxies where the true DM distribution is known. In these controlled environment, the tool successfully discards the NFW  model, a result that aligns  with the actual DM mass distributions computed within the simulations. Plummer-Schuster and $\rho_{230}$ potentials are not discarded. This validation is further reinforced by a successful test on a FIRE simulation low-mass halo in Appendix~\ref{app:fire}, confirming that the tool maintains physical consistency even in realistic numerical settings.

In summary, \eimt\ has been  validated in a diverse range of systems. By assessing physical validity for the stellar distribution function, the tool distinguished cored and cuspy profiles, correctly identifying the cored nature of most globular clusters and dSphs like Fornax while accurately allowing a cuspy profile for Draco. This broad agreement with established kinematic studies and simulation ground truth confirms \eimt\ as a robust and efficient diagnostic for constraining possible shapes of DM halos using only photometric data.


\begin{acknowledgements}
We are most grateful to Ignacio Trujillo for accompanying the development of this work with helpful suggestions and insightful comments.
Thanks are also due to  Claudio Dalla Vecchia for support and help with the numerical aspects of the work.
The suggestions of an anonymous referee contributed to clarify some of the passages of the paper.
The analysis of the GCs was originally started with the master thesis of Javier Moya Plaza.
We are particularly grateful to the researchers who provided the various datasets analyzed in this work: Barbara Lanzoni and Francesco F. Ferraro for the GC data (Sect.~\ref{sec:globular}); José María Arroyo-Polonio for the Sculptor data (Sect.~\ref{sec:sculptor}); Giuseppina Battaglia for the Draco data (Sect.~\ref{sec:draco}); and Justin Read and Fernando Valenciano for the EDGE halo data (Sect.~\ref{sec:edge}).
The research is partly  funded through grant PID2022-136598NB-C31 (ESTALLIDOS 8) by the Spanish Ministry of Science and Innovation (MCIN/AEI/10.13039/501100011033)  and {\em ERDF A way of making Europe}.
The author has been supported by the European Union through the grant UNDARK' of the widening participation and spreading excellence programe (project number 101159929).
The authors acknowledge the use of ChatGPT (OpenAI) for assistance with English language. All scientific interpretations, analyses, and conclusions are those of the authors.
\end{acknowledgements}

%

\bibliography{biblio_paper170.bib}{}
\bibliographystyle{aa}

\appendix
\section{Test using a Feedback In Realistic Environments (FIRE) numerically simulated halo}\label{app:fire}
\begin{figure}
\centering
\includegraphics[width=0.85\columnwidth]{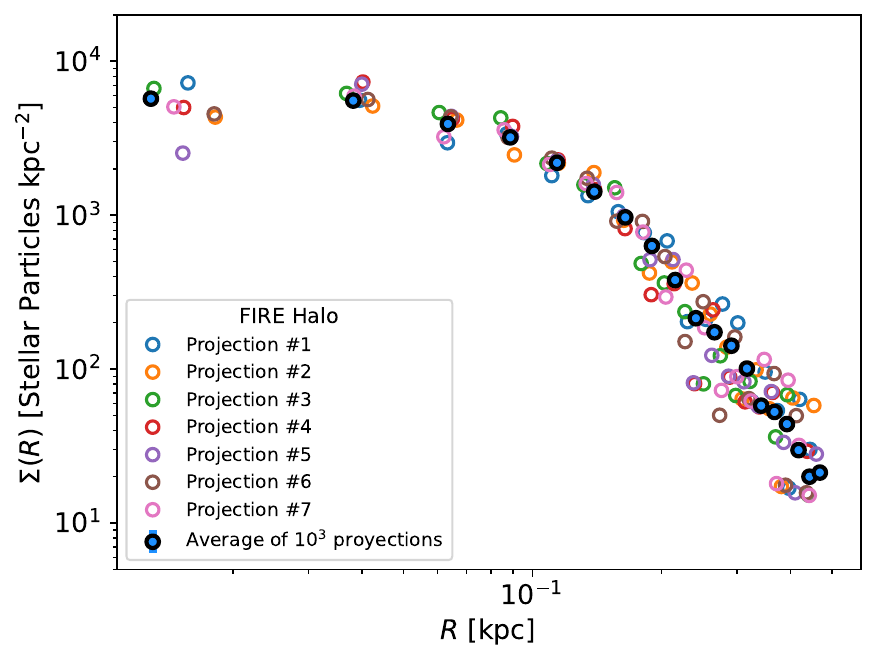}
\caption{
Stellar surface density of the model low-mass FIRE-2 halo described in Appendix~\ref{app:fire}. One thousands 2D projections of the stellar distribution are averaged in rings to produce the radial profile shown as full blue symbols.  The open symbols of other colors show seven different random projections. 
}
\label{fig:profile}
\end{figure}

During the early phases of the development of \eimt, while it was being applied by \citet{2024ApJ...973L..15S} to UFDs, Nitya Kallivayalil
and Jack Warfield \citep[authors of the paper from which the UFD profiles were taken;][]{2024ApJ...967...72R} suggested testing the method on a faint halo from the FIRE-2 suite of numerical simulations \citep{2018MNRAS.480..800H}. A galaxy ({\tt m09\_res30\_gal23512})  was selected from the high resolution run \citep{2019MNRAS.490.4447W, 2019MNRAS.489.4574G}, with  
a DM halo mass $\sim\,10^9\,M_\odot$ and a stellar mass as low as $\sim\,10^4\,M_\odot$.
The average of a 1000 randomly oriented 2D projections of the stellar distribution were used 
to compute the average profile shown in Fig.~\ref{fig:profile} (the blue solid symbols). The radial profiles were computed averaging in 20 rings, which assumes the system to be spherically symmetric. This approximation seems reasonable, as illustrated in Fig.~\ref{fig:profile}: if the system were significantly triaxial, the individual projections would differ in extent, which is not the case (cf. the open symbols and the average profile).

\begin{figure}
\centering
\includegraphics[width=0.9\linewidth]{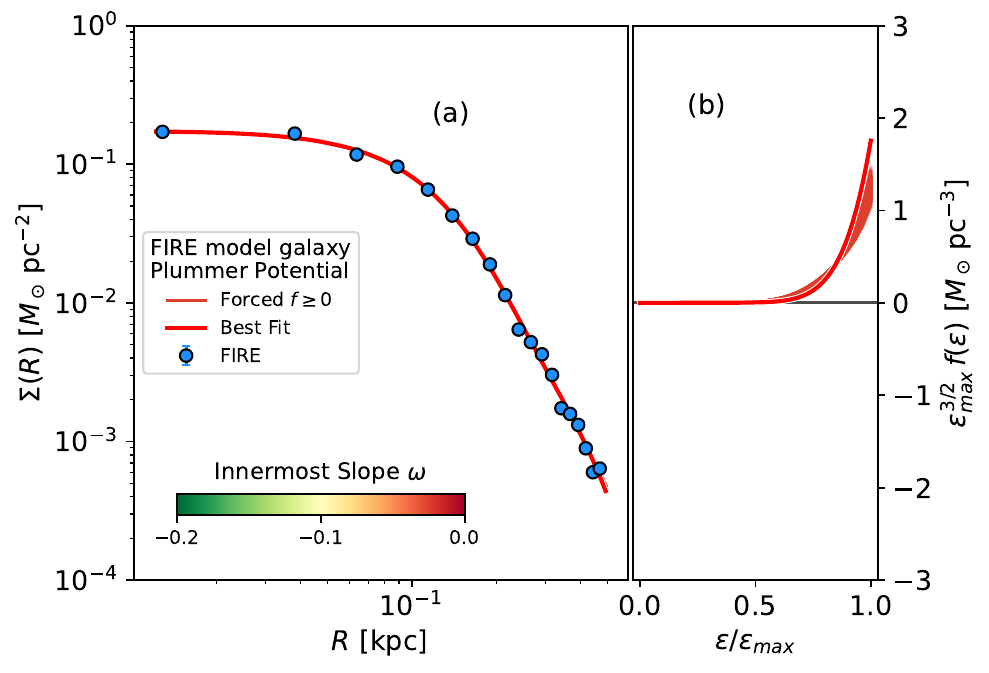}\\
\includegraphics[width=0.9\linewidth]{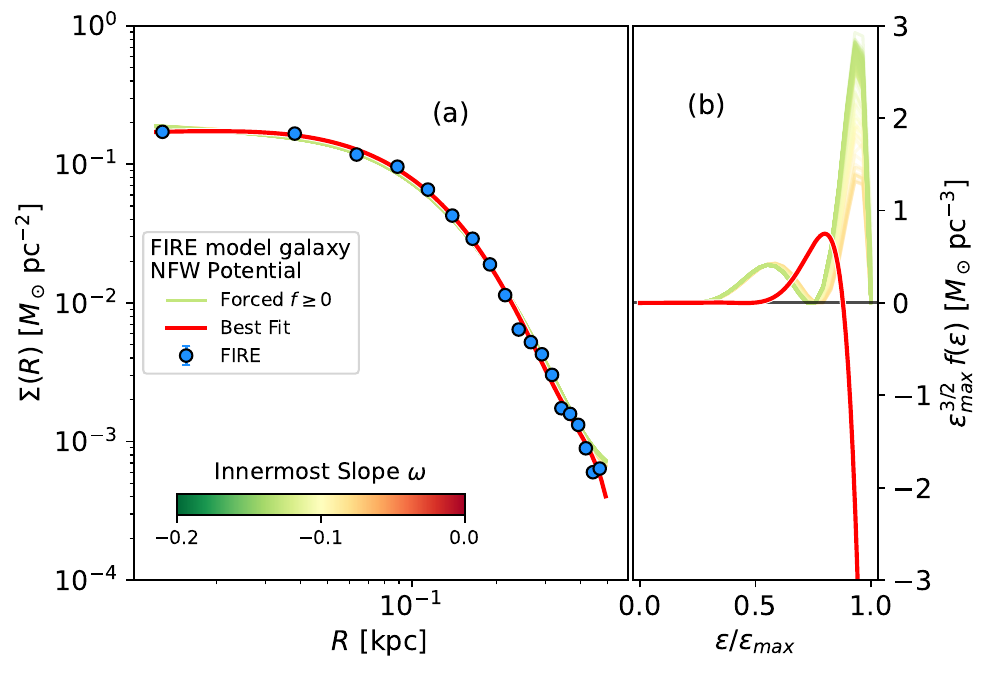}\\
\includegraphics[width=0.80\linewidth]{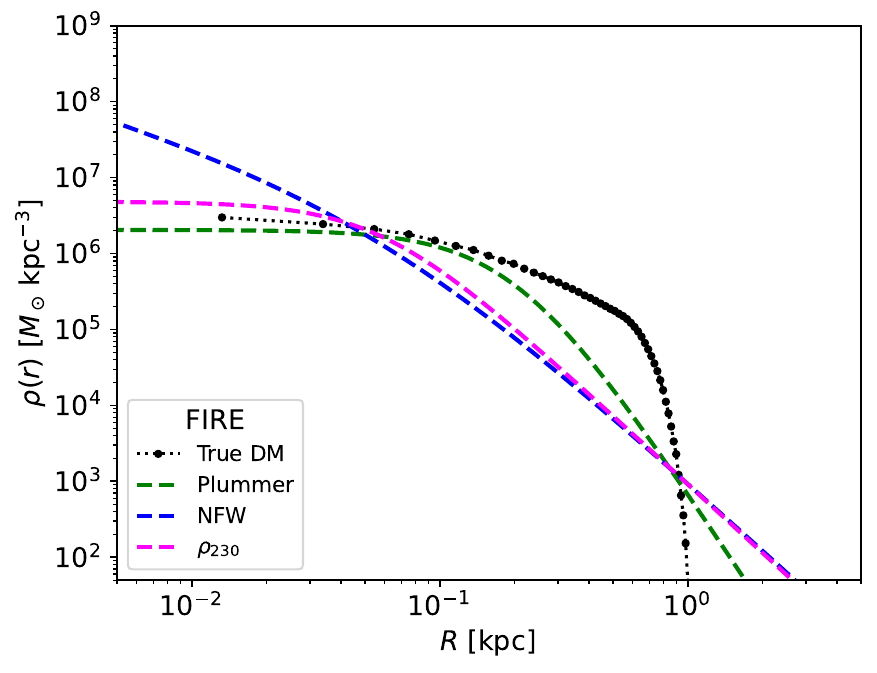}
\caption{
Fits corresponding to the surface density of a simulated low mass galaxy from FIRE-2, as described in Appendix~\ref{app:fire}. 
Top panel: Fit and $f$ when a Schuster-Plummer potential is assumed.  Middle panel: same as the top panel but assuming the potential to be NFW, which has a cusp and whose best fit requires $f <0$ (the solid red lines). Bottom panel: True 3D DM distribution (black symbols), which should be compared with the best-fitting distributions corresponding to the three different potentials for $f \ge 0$ (the dashed lines, as labelled in the inset). Evidently, the true DM distribution deviates from the cuspy NFW profile that  best-fits the stellar surface density. If anything, the true potential exhibits a core, although it is more extended than those of the best-fitting models. 
}
\label{fig:df6_run_plot_601_pub}
\end{figure}

The result of applying the \eimt\ to the average stellar surface density is shown in Fig.~\ref{fig:df6_run_plot_601_pub}. 
A Plummer potential provides a good fit with a physically sensible $f\geq 0$ (the red solid line in  Fig.~\ref{fig:df6_run_plot_601_pub}, top panel).  This agreement contrasts with the result assuming an NFW potential for which  the best fit demands $f < 0$ (Fig.~\ref{fig:df6_run_plot_601_pub}, middle panel), indicating that  the model galaxy is not living in an NFW halo. The fit assuming a $\rho_{230}$ potential (not shown) is also physically consistent since it does not prefer a negative distribution function.

Thus, the application of \eimt\ to this object leads to the conclusion that the stars are not hold together by an NFW potential, however, they can stay in a cored potential (Schuster-Plummer or $\rho_{230}$). The  DM distributions best fitting the stars, provided $f \geq 0$, are shown in the lower panel of  Fig.~\ref{fig:df6_run_plot_601_pub}. The figure also includes the true DM distribution (black symbols) which, in agreement with the conclusion drawn from \eimt ,  is very different from an NFW halo. Moreover, the true DM distribution has a core as the DM distributions allowed by \eimt , although the outskirts of the true DM profile does not look like  a Schuster-Plummer or a $\rho_{230}$ DM distribution.

We note that the inner core of both the stars and DM (Fig.~\ref{fig:df6_run_plot_601_pub}) are artificially created in the simulation by the use of a finite softening length of $\sim$\,40\,pc. Thus,  artificial collision have thermalized the DM distribution  preventing the model galaxy to show the cuspy DM distribution expected in CDM \citep{2021MNRAS.504.2832S}. However, this fact should not pose a problem for applying \eimt\  which does not care about how the balance between the location of the stars and the potential is stablished, provided an stationary state is reached. 

%
%
%

%

\section{The observed stellar surface density profile for Draco}\label{app:draco}

\begin{figure}
\centering
\includegraphics[width=0.9\linewidth]{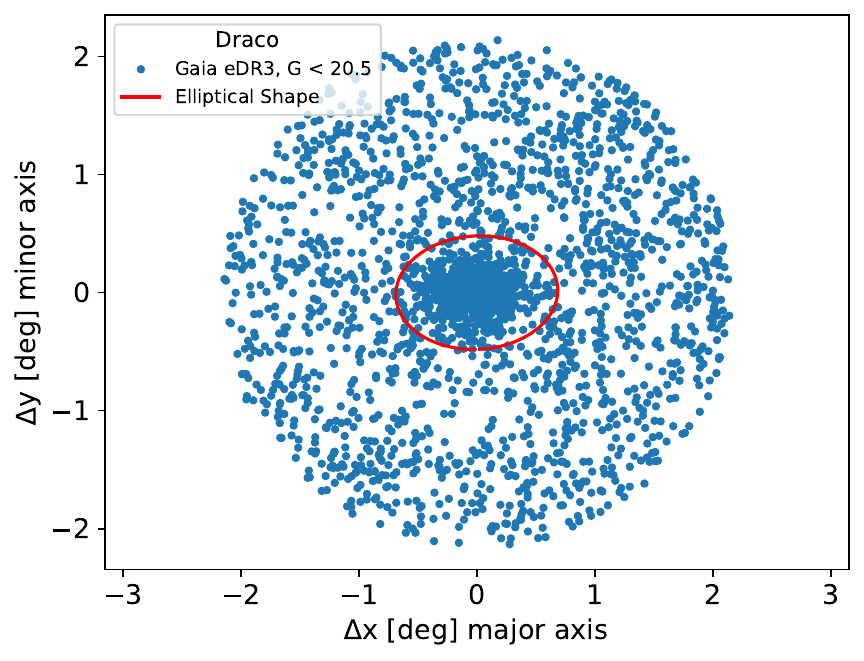}\\
\includegraphics[width=0.9\linewidth]{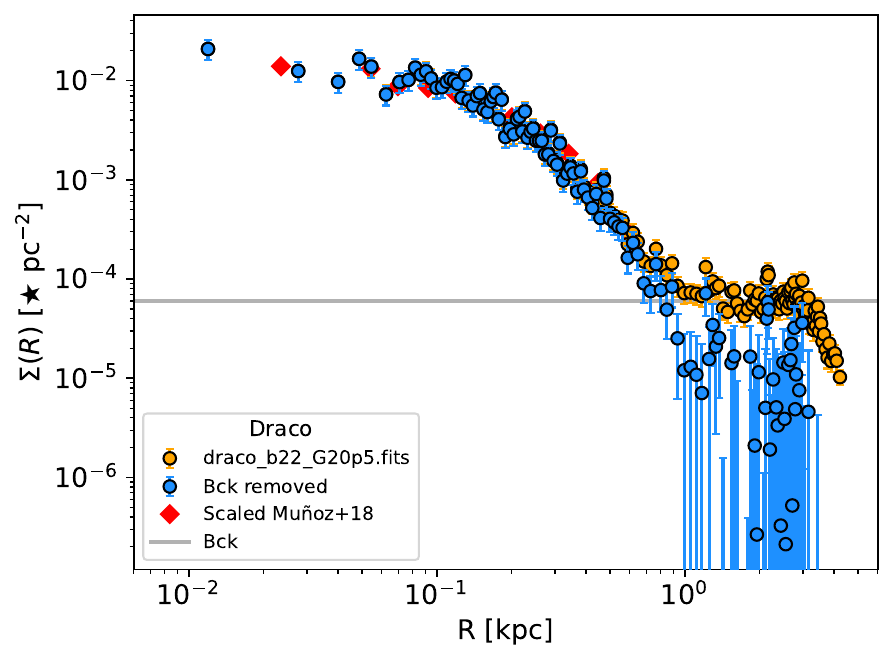}
\caption{
Constructing the observed surface density profile of Draco.
Top panel: $x$  and $y$ on the sky, with $\Delta$ implying coordinates with respect to the center of the galaxy. The red line represents the ellipticity and PA taken \citet{2018ApJ...860...66M}.   Bottom panel: surface density profile derived from the top panel in bins with equal number of stars. The background is marked as a dashed black line and has been found by trial and error. The red symbols represent previous data for this galaxy from  \citet{2018ApJ...860...66M}.
}
\label{fig:read_battaglia_1}
\end{figure}

%
The profile of Draco used in our tests come from stellar counts form Gaia eDR3 \citep{2021A&A...649A...1G}. Sources are extracted over regions extending beyond 5 half-light radii from the center of Draco and filtered using stringent astrometric and photometric quality cuts, excluding duplicates, extended objects, and AGN matches. For details, we refer to Sect.~3 of  \citet{2022A&A...657A..54B}. We employ two different absolute magnitude cuts, $G < 20\,{\rm mag}$ and $G < 20.5\,{\rm mag}$, but both giving the same consistent results. 
The structural parameters (center, position angle PA, and ellipticity) and the distance were taken from the literature  \citep{2018ApJ...860...66M,2022A&A...657A..54B}. Using them, AR and DEC  were transformed to $x$ and $y$ in a reference system on the sky with $x$ along the major axis (Fig.~\ref{fig:read_battaglia_1}, top panel). The minor axes are scaled up to the mayor axes according to the ellipticity and, using these coordinates, the distance to the center of each star was computed. Thus, all the stars over an ellipse of the assumed PA and ellipticity have the same distance, which coincides with the mayor axis of the ellipse. To ensure equal counts per bin, stars are sorted by radial distance in these coordinates and grouped into bins containing the same number of objects. We take 20 stars/bin, but the profiles do not change significantly upon changing this selection. The result is shown in Fig.~\ref{fig:read_battaglia_1}, bottom panel, with the error bars accounting for the Poisson statistics when counting. The radius in the figure represents the major axes of the elliptical annuli, but the areas to compute densities are true areas projected on the sky. The subtraction of a constant background (the dashed line in Fig.~\ref{fig:read_battaglia_1}) only affect the outskirts, and after this manipulation, one finally obtains the data used in our analysis.   Except for a scaling factor explained by the different depth of the observation, the resulting profile is very much like the profile derived for this galaxy by \citet{2018ApJ...860...66M}, the latter included  as red diamonds in Fig.~\ref{fig:read_battaglia_1}, the bottom panel.

The fits of the Draco surface density carried out in Sect.~\ref{sec:draco} assume a mass-to-light ratio of one in solar units. As we explain in Sec.~\ref{sec:briefing}, the actual value is not relevant since the fits are independent of a global scaling in ether the stellar mass or the DM halo mass. 


\end{document}